\documentclass[onecolumn,11pt]{article}
\usepackage[table]{xcolor}
\usepackage{amsmath,amsfonts,amscd,amssymb}
\usepackage{graphicx}
\usepackage{epstopdf}
\usepackage{psfrag}
\usepackage{overpic}
\usepackage{cancel}
\usepackage{rotating}
\usepackage{url}
\usepackage{caption}
\usepackage{color}
\usepackage{rotating}
\usepackage{multirow}
\usepackage{wrapfig}
\usepackage{mathtools}
\usepackage{subeqnarray}
\usepackage{setspace}
\usepackage{palatino} % needs 10
\usepackage[numbers,sort&compress]{natbib}
\usepackage{booktabs}

\usepackage{array}
\newcolumntype{L}{>{\centering\arraybackslash}m{10cm}}
\usepackage{authblk}

\newcommand\blfootnote[1]{%
  \begingroup
  \renewcommand\thefootnote{}\footnote{#1}%
  \addtocounter{footnote}{-1}%
  \endgroup
}

\definecolor{header1}{cmyk}{0,0,0,1}

\usepackage{amsthm}

\DeclareMathOperator*{\argmin}{arg\rm{}min}

\newcommand{\bL}{\boldsymbol{L}}
\newcommand{\bI}{\boldsymbol{I}}
\newcommand{\balpha}{\boldsymbol{\alpha}}
\newcommand{\bA}{\boldsymbol{A}}

\newcommand{\bX}{\boldsymbol{X}}

\newcommand{\bY}{\boldsymbol{Y}}
\newcommand{\bF}{\boldsymbol{F}}
\newcommand{\bG}{\boldsymbol{G}}
\newcommand{\bW}{\boldsymbol{W}}

\newcommand{\chol}{\boldsymbol{C}}
\newcommand{\bN}{\boldsymbol{N}}

\newcommand{\bM}{\boldsymbol{M}}
\newcommand{\bV}{\boldsymbol{V}}
\newcommand{\bs}{\boldsymbol{s}}
\newcommand{\bS}{\boldsymbol{S}}
\newcommand{\bomega}{\boldsymbol{\omega}}
\newcommand{\bOmega}{\boldsymbol{\Omega}}
\newcommand{\bSigma}{\boldsymbol{\Sigma}}
\newcommand{\bU}{\boldsymbol{U}}
\newcommand{\bb}{\boldsymbol{b}}
\newcommand{\bfun}{\boldsymbol{f}}
\newcommand{\bg}{\boldsymbol{g}}
\newcommand{\bl}{\boldsymbol{l}}
\newcommand{\bx}{\boldsymbol{x}}
\newcommand{\bc}{\boldsymbol{c}}
\newcommand{\bLambda}{\boldsymbol{\Lambda}}
\newcommand{\ba}{\boldsymbol{a}}

\newcommand{\bXi}{\boldsymbol{\Xi}}
\newcommand{\bPi}{\boldsymbol{\Pi}}
\newcommand{\bpi}{\boldsymbol{\pi}}

\newcommand{\bh}{\boldsymbol{h}}

\newcommand{\bP}{\boldsymbol{P}}
\newcommand{\bPhi}{\boldsymbol{\Phi}}
\newcommand{\bPsi}{\boldsymbol{\Psi}}
\newcommand{\bpsi}{\boldsymbol{\psi}}
\newcommand{\bphi}{\boldsymbol{\phi}}
\newcommand{\by}{\boldsymbol{y}}
\newcommand{\bv}{\boldsymbol{v}}
\newcommand{\bu}{\boldsymbol{u}}
\newcommand{\bw}{\boldsymbol{w}}

\newcommand{\bTheta}{\boldsymbol{\Theta}}

\newcommand{\btheta}{\boldsymbol{\theta}}
\newcommand{\bxi}{\boldsymbol{\xi}}
\newcommand{\bk}{\boldsymbol{k}}
\newcommand{\bD}{\boldsymbol{D}}
\newcommand{\bK}{\boldsymbol{K}}
\newcommand{\bUp}{\boldsymbol{U}}
\newcommand{\bVp}{\boldsymbol{V}}
\newcommand{\bSp}{\boldsymbol{\Sigma}}

\renewcommand{\H}{\mathcal{H}}

\usepackage{amssymb}
\usepackage{lipsum}
\usepackage{tikz,pgfplots,pstool}

\pgfplotsset{compat=1.13}

\usepackage{contour}
\usepackage[normalem]{ulem}

\contourlength{0.8pt}

\newcommand{\myuline}[1]{%
  \uline{\phantom{#1}}%
  \llap{\contour{white}{#1}}%
}

\usepackage{subcaption}

\newtheoremstyle{mylemmastyle}
  {3pt} % Space above
  {3pt} % Space below
  {\itshape} % Body font
  {} % Indent amount
  {\bfseries} % Theorem head font
  {.} % Punctuation after theorem head
  {\newline} % Space after theorem head
  {\thmname{#1} \thmnumber{#2}: \thmnote{\normalfont\myuline{#3}}} % Theorem head spec (can be left empty, meaning `normal')
\theoremstyle{mylemmastyle} 
\newtheorem{mylemma}{Lemma}

\usepackage{algorithm}
\usepackage{algorithmicx}
\usepackage{algpseudocode}
\algrenewcommand{\algorithmiccomment}[1]{\hfill #1}

\usepackage{hyperref}

\hypersetup{
    colorlinks,
    linkcolor={red!70!black},
    citecolor={blue!70!black},
    urlcolor={green!50!black}
}
\definecolor{col0}{RGB}{82,173,2}
\definecolor{col1}{RGB}{212,173,2}
\definecolor{col2}{RGB}{255, 111, 0}
\usepackage[bottom,flushmargin,hang,multiple]{footmisc}
\usepackage[top=1in, bottom=1in, left=1in, right=1in]{geometry}

\title{\vspace{-.45in} \fontsize{17}{17}{\textbf{Kernel Learning for Robust Dynamic Mode Decomposition:}}\\\fontsize{16.5}{16.5}{\textbf{\hspace{-.04in}Linear and Nonlinear Disambiguation Optimization (LANDO)}}\vspace{-.1in}}
\author[1]{%
Peter J. Baddoo$^*$}
\author[2,3]{Benjamin Herrmann}
\author[4]{Beverley J. McKeon}
\author[2]{Steven L. Brunton\vspace{-.1in}}

\affil[1]{\small Department of Mathematics, Massachusetts Institute of Technology,
 Cambridge, MA 02139, USA}
\affil[2]{Department of Mechanical Engineering, University of Washington, Seattle, WA 98195, USA}
\affil[3]{Institute of Fluid Mechanics, Technische Universit\"at Braunschweig, 38108 Braunschweig, Germany}
\affil[4]{Graduate Aerospace Laboratories, California Institute of Technology, Pasadena CA 91125, USA}

\date{}

\begin{document}
\maketitle
\blfootnote{$^*$ Corresponding author (baddoo@mit.edu)}
\vspace{-.4in}
\begin{abstract}
Research in modern data-driven dynamical systems is typically focused on the three key challenges of high dimensionality, unknown dynamics, and nonlinearity.
The dynamic mode decomposition (DMD) has emerged as a cornerstone for modeling high-dimensional systems from data. 
However, the quality of the linear DMD model is known to be fragile with respect to strong nonlinearity, which contaminates the model estimate.  
In contrast, sparse identification of nonlinear dynamics (SINDy) learns fully nonlinear models, disambiguating the linear and nonlinear effects, but is restricted to low-dimensional systems. 
In this work, we present a kernel method that learns interpretable data-driven models for high-dimensional, nonlinear systems.
Our method performs kernel regression on a sparse dictionary of samples that appreciably contribute to the underlying dynamics.
We show that this kernel method efficiently handles high-dimensional data and is flexible enough to incorporate partial knowledge of system physics. 
It is possible to accurately recover the linear model contribution with this approach, disambiguating the effects of the implicitly defined nonlinear terms, resulting in a DMD-like model that is robust to strongly nonlinear dynamics. 
We demonstrate our approach on data from a wide range of nonlinear ordinary and partial differential equations that arise in the physical sciences. 
This framework can be used for many practical engineering tasks such as model order reduction, diagnostics, prediction, control, and discovery of governing laws.
\end{abstract}

\section{Introduction}
\label{sec:intro}
Discovering interpretable patterns and models from high-dimensional data is one of the principal challenges of scientific machine learning, with the potential to transform our ability to predict and control complex physical systems~\cite{Brunton2019}. 
The current surge in the quality and quantity of data, along with rapidly improving computational hardware, has motivated a wealth of machine learning techniques that uncover such patterns for dynamical systems. 
Successful recent methods include the dynamic mode decomposition (DMD)~\cite{Schmid2010,Rowley2009,Tu2014,Kutz2016} and extended DMD~\cite{Williams2015a,Williams2015b}, sparse identification of nonlinear dynamics (SINDy) for ordinary and partial differential equations~\cite{Brunton2016,Rudy2017}, genetic programming for model discovery~\cite{Bongard2007pnas,Schmidt2009science}, physics informed neural networks (PINNs)~\cite{Raissi2019,Raissi2020}, Lagrangian neural networks~\cite{Cranmer2020}, time-lagged autoencoders~\cite{Wehmeyer2018}, operator theoretic methods~\cite{Mezic2013,Kaiser2014jfm,Klus2018,Brunton2021}, and operator inference~\cite{Peherstorfer2016,Qian2020}. 
Techniques based on generalized linear regression, such as DMD and SINDy, are widely used because they are computationally efficient, require less data than neural networks, are highly extensible, and provide interpretable models.   
However, these approaches are either challenged by nonlinearity (e.g., DMD) or don't scale to high-dimensional systems (e.g., SINDy). 
In this work, we present a machine learning algorithm that leverages sparse kernel regression to address both challenges, efficiently learning high-dimensional nonlinear models that admit interpretable spatio-temporal coherent structures and robust locally linear models.

\begin{figure}[t]
\vspace{-.15in}
	\centering
	\begin{overpic}[width = .75\linewidth]{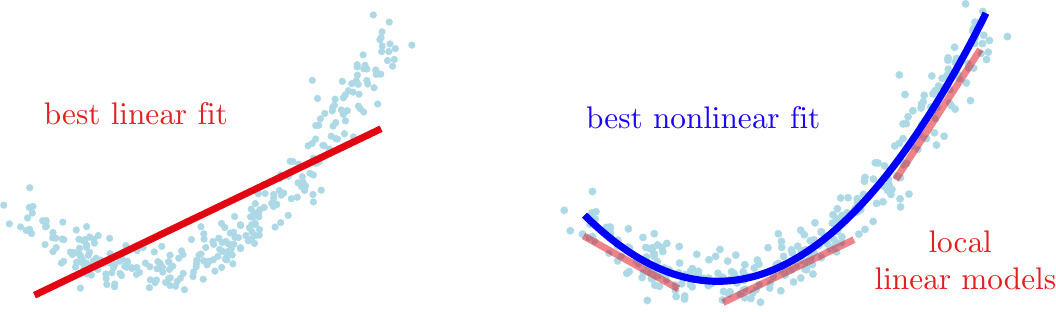}
	\put(-2,27){(a)}
	\put(51,27){(b)}
	\end{overpic}
	\vspace{-.1in}
	\caption{Learning regression models in linear (a) and nonlinear (b) feature spaces. Our approach disambiguates linear and nonlinear model contributions to accurately extract local linear models. 
}
\label{Fig:linearisedPic}
\vspace{-.15in}
\end{figure}

A central goal of modern data-driven dynamical systems~\cite{Brunton2019} is to identify a model 
\begin{align}\label{Eq:ContinuousDynamics}
    \frac{d}{dt}{\bx} = {\bF}(\bx) = \bL \bx + \bN(\bx)
\end{align}
that describes the evolution of the state of the system, $\bx$. 
Here we explicitly indicate that the dynamics $\bF$ have a linear $\bL$ and nonlinear $\bN$ contribution, although many techniques do not model these separately or explicitly.  
However, several approaches obtain interpretable and explicit models of this form.  
For example, DMD seeks a best-fit linear model of the dynamics, while SINDy directly identifies sparse nonlinear models of the form in \eqref{Eq:ContinuousDynamics}. 

Our approach synthesizes favorable aspects of several approaches mentioned above; however, it most directly complements and addresses the challenges of DMD for strongly nonlinear systems. 
The dynamic mode decomposition was originally introduced by Schmid~\cite{Schmid2010} in the fluid dynamics community as a method for extracting spatio-temporal coherent structures from high-dimensional data, resulting in a low-rank representation of the best-fit linear operator that maps the data forward in time~\cite{Tu2014,Kutz2016}. 
The resulting linear DMD models have been used to characterize many systems in fluid mechanics, where complex flows admit dominant modal decompositions~\cite{Lumley:1970,taira2017aiaa,Towne2018,taira2019aiaa} and linear control is commonly used~\cite{fabbiane2014amr,brunton2015amr,sipp2016amr,rowley2017arfm}. 
DMD has also been adopted in a wide range of fields beyond fluid mechanics, and much of its success stems from the formulation of DMD as a linear regression problem~\cite{Tu2014}, based entirely on measurement data, resulting in several powerful extensions~\cite{Kutz2016}.  
However, because DMD uses least-squares regression to find a best-fit linear model $d\bx/dt\approx\bA\bx$ to the data, the presence of measurement noise~\cite{Bagheri2014pof,Dawson2016,Hemati2017tcfd,Askham2018}, control inputs~\cite{Proctor2016}, and nonlinearity bias the regression. 
Mathematically, the noise, control inputs, and nonlinearity may all be lumped into a forcing $\bb$:
\begin{align}\label{eq:DMDcontaminated}
    \frac{d}{dt}\bx = \bL\bx + \bb \approx \bA\bx.
\end{align}
The forcing $\bb$ contaminates the linear model estimate, so $\bA$ from DMD does not approximate the true linear contribution from $\bL$.
It was recognized early that the DMD algorithm was highly sensitive to noise~\cite{Bagheri2014pof,Dawson2016,Hemati2017tcfd}, resulting in noise-robust variants, including forward backward DMD~\cite{Dawson2016}, total least-squares DMD~\cite{Hemati2017tcfd}, optimized DMD~\cite{Askham2018}, consistent DMD~\cite{azencot2019consistent}, and DMD based on robust PCA~\cite{Scherl2020prf}.  
Similarly, DMD with control (DMDc)~\cite{Proctor2016} was introduced to disambiguate the effect of the linear dynamics from actuation.  
For statistically stationary systems with stochastic inputs, the spectral proper orthogonal decomposition (SPOD)~\cite{Lumley:1970} produces an optimal basis of modes to describe the variability in an ensemble of DMD modes~\cite{Towne2018}.
The bias due to nonlinearity, shown in Fig.~\ref{Fig:linearisedPic}(a), has been less thoroughly explored and is the topic of the present work.  

Despite these challenges, DMD is frequently applied to strongly nonlinear systems, with theoretical motivation from Koopman operator theory~\cite{Rowley2009,Mezic2013,Kutz2016,Brunton2021}. 
However, DMD typically only yields accurate models for periodic or quasiperiodic systems, and is fundamentally unable to capture nonlinear transients, multiple fixed points or periodic orbits, or other more complicated attractors~\cite{Brunton2016plosone}. 
Williams et al.~\cite{Williams2015a} developed the \emph{extended DMD} (eDMD), which augments the original state with nonlinear functions of the state to better approximate the nonlinear eigenfunctions of the Koopman operator for nonlinear systems. 
Further, a kernel version of eDMD was introduced for high-dimensional systems~\cite{Williams2015b}. 
However, because this approach still fundamentally results in a linear model (in the augmented state), it also suffers from the same issues of not being able to handle multiple fixed points or attracting structures, and it also typically suffers from closure issues related to the irrepresentability of Koopman eigenfunctions. 
The sparse identification of nonlinear dynamics~\cite{Brunton2016} algorithm is a related regression approach to model discovery, which identifies a fully nonlinear model as a sparse linear combination of candidate terms in a library. 
While SINDy is able to effectively disambiguate the linear and nonlinear dynamics in \eqref{Eq:ContinuousDynamics}, resulting in the ability to obtain de-biased locally linear models as in Fig.~\ref{Fig:linearisedPic}(b), it only applies to relatively low-dimensional systems because of poor scaling of the library with state dimension. 

\subsection{Contributions of this work}
\begin{figure}[t]
\vspace{-.15in}
	\centering
	\includegraphics[width=.99\linewidth]{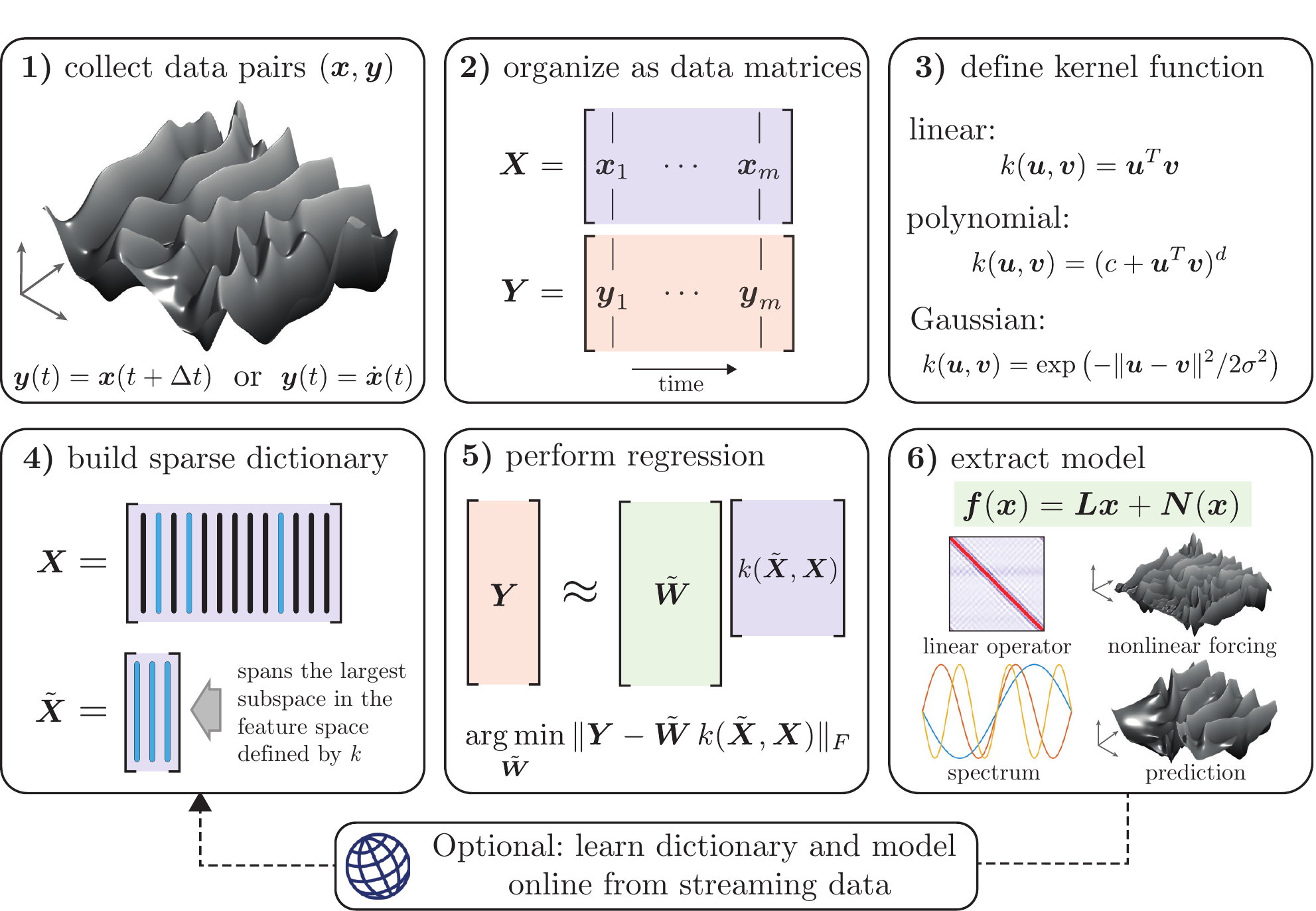}
	\caption{The Linear and Nonlinear Disambiguation Optimization (LANDO) framework.
	Training data in the form of snapshot pairs 
	are collected from either simulation or experiment in 1).
The data are organised into data matrices in 2).
In 3) an appropriate kernel is defined, which can be informed by
expert knowledge of the underlying physics of the system or
through a cross-validation procedure.
In 4) a sparse dictionary of basis elements is constructed 
from the training samples, and in 5) the regression problem is solved.
Finally, in 6) an interpretable model is extracted.
	}%
	\label{Fig:main-diagram}
	\vspace{-.1in}
\end{figure}

In this work, we develop a custom kernel regression algorithm to learn accurate, efficient, and interpretable data-driven models for strongly nonlinear, high-dimensional dynamical systems.  
This approach scales to very high-dimensions, unlike SINDy, yet still accurately disambiguates the linear part of the model from the implicitly defined nonlinear dynamics. 
Thus, it is possible to obtain linear DMD models, local to a given base state, that are robust to strongly nonlinear dynamics.  
Our approach, referred to as the linear and nonlinear disambiguation optimization (LANDO) algorithm, may be viewed as a generalisation of DMD that enables a robust disambiguation of the underlying linear operator from nonlinear forcings.
The learning framework is illustrated in Fig.~\ref{Fig:main-diagram}, and open-source code is available at \url{www.github.com/baddoo/LANDO}.

To achieve this robust learning, we improve upon several leading kernel and system identification algorithms.  
Recent works have successfully applied kernel methods~\cite{Scholkopf2002,Shawe-Taylor2004} to study data-driven dynamical systems~\cite{Williams2015b,Bouvrie,Kawahara2016,Li2017chaos,Owhadi2019,Giannakis2019acha,das2020koopman,Gelss2020Kernel,li2020neural,manohar2020kernel}. 
A key contribution, and an inspiration for the present work, is kernel DMD (kDMD,~\cite{Williams2015b}), which seeks to approximate the infinite-dimensional Koopman operator as a large square matrix evolving nonlinear functions of the original state.  
An essential difference between kDMD and the present work is that our goal is to implicitly model the (non-square) nonlinear dynamics in \eqref{Eq:ContinuousDynamics} in terms of the original state $\bx$,  enabling the robust extraction of the linear component $\bL$, as opposed to analysing the Koopman operator over measurement functions.
In our work, we present a modified kernel recursive least-squares algorithm (KRLS,~\cite{Engel2004}) to learn a nonlinear model that best characterizes the observed data.  
To reduce the training cost, which typically scales with the cube of the number of training samples for kernel methods, we use dictionary learning to iteratively identify samples that contribute to the dynamics. 
This dictionary learning approach may be seen as a sparsity promoting regulariser, significantly reducing the high condition number that is common with kernel methods, improving robustness to noise. 
We introduce an iterative Cholesky update to construct the dictionary in a numerically stable manner, significantly reducing the training cost while also mitigating overfitting. 
Similarly to KRLS, our model has the option of operating online by parsing data in a streaming fashion and exploiting rank-one matrix updates to revise the model.
Therefore, our approach is also suitable for model order reduction in practical applications where data become available ''on-the-fly``. 
Further, we show how to incorporate partially known physics, or uncover unknown physics, by designing or testing tailored kernels, much as with the SINDy framework~\cite{Brunton2016,Loiseau2018}.

We demonstrate our proposed kernel learning approach on a range of complex dynamical systems that arise in the physical sciences.  
As an illustrative example, we first explore the chaotic Lorenz system. 
We also consider partial differential equations, using the LANDO framework to uncover the linear and nonlinear components of the nonlinear Burgers', Korteweg-De Vries, and Kuramoto--Sivashinsky equations using only nonlinear measurement data. 
The algorithm accurately recovers the spectrum of the linear operator for these systems, enabling linearised analyses, such as linear stability, transient growth, and resolvent analyses~\cite{jovanovic2005jfm,Mckeon2010,Mckeon2017,jovanovic2021arfm}, in a purely data-driven manner~\cite{herrmann2021jfm}, even for strongly nonlinear systems.
We also demonstrate the approach on a high-dimensional system of coupled nonlinear Kuramoto oscillators. 

The remainder of the work is organized as follows.  
Section~\ref{Sec:mathematicalBackground} provides a mathematical background overview of the DMD and kernel methods. 
Section~\ref{Sec:method} introduces our kernel learning procedure for dynamical systems, including the sparse dictionary learning with Cholesky updates. 
We demonstrate how to extract interpretable structures from these kernel models, such as robust linear DMD models, in Section~\ref{Sec:ExtractingStructure}. 
Results on a variety of nonlinear dynamical systems are presented in Section~\ref{Sec:Applications}.  
Finally, Section~\ref{Sec:Discussion} concludes with a discussion of limitations and suggested extensions of the method, providing a comparison with the related DMD, eDMD/kDMD, and SINDy approaches, also summarised in Fig.~\ref{Fig:methodsComparison}. 
In the appendices we explicate the connection between our method and DMD, demonstrate how one can incorporate the effects of control, present the equations for online updating, and investigate the sensitivity of the algorithm to noise.  
\section{Problem statement and mathematical background}
\label{Sec:mathematicalBackground}
In this section we will define our machine learning problem and 
review some relevant mathematical ideas related to DMD (Sec.~\ref{Sec:exactDMD}) and kernel methods (Sec.~\ref{Sec:kernel}). 

We consider dynamical systems describing the evolution of an $n$-dimensional vector ${\bx\in\mathbb{R}^n}$ that characterizes the state of the system.  
We will consider both continuous-time and discrete-time dynamics in this work. 
The dynamics may be expressed either in continuous time as
\begin{align*}
    \frac{d}{dt}{\bx}(t) = \bF(\bx(t))
\end{align*}
or in discrete time as
\begin{align*}
    \bx_{j+1} = \bF(\bx_j).
\end{align*}
For a given physical system, the continuous-time and discrete-time representations of the dynamics will correspond to different functions $\bF$, although we use the same function above for notational simplicity. 
In general, the dynamics may also vary in time and depend on control inputs $\bu$ and parameters $\boldsymbol{\beta}$; however, for simplicity, we begin with the autonomous dynamics above. 

Our goal is to learn a tractable representation of the dynamical system ${\bF: \mathbb{R}^n \rightarrow \mathbb{R}^n}$ that is both accurate and interpretable, informing tasks such as physical understanding, diagnostics, prediction, and control.
We suppose that we have access to a training set of data pairs $\{(\bx_j,\by_j)\in \mathbb{R}^n \times \mathbb{R}^n \,|\, j = 1,\dots, m\}$,
which are connected through the dynamics by
\begin{align}
	\by_j = \bF(\bx_j).% \qquad \textrm{for } j = 1, \dots, m.
\end{align}
If the dynamics continuous in  time, $\by_j=\dot{\bx}_j$, where the dot denotes differentiation in time, and if the dynamics are expressed in discrete time $\by_j = \bx_{j+1}$. 
The discrete-time formulation is more common, as data from simulations and experiments is often sampled or generated at a fixed sampling interval $\Delta t$, so $\bx_j=\bx(j\Delta t)$. 
However, this work applies to both discrete and continuous systems, and the only practical difference arises in the eigenvalues of linearized models.  

The training data correspond to $m$ snapshots in time of a simulation or experiment. 
For ease of notation, it is typical to arrange the samples into snapshot data matrices of the form
\begin{align}\label{Eq:SnapshotMatrices}
\boldsymbol{X}=  \begin{bmatrix} | & | &  & | \\ \bx_1 & \bx_2 & \cdots & \bx_{m} \\ | & | &  & |\end{bmatrix},
\qquad
\boldsymbol{Y} =  \begin{bmatrix} | & | &  & | \\ \by_1 & \by_2 & \cdots & \by_{m} \\ | & | &  & | \end{bmatrix}.
\end{align} 
In many applications the state dimension is much larger than the number of snapshots, so $m \ll n$.
For example, the state may correspond to a fluid velocity field sampled on a discretized grid. 

Our machine learning problem consists of finding a function $\bfun$ that suitably maps the training data given certain generalisability, interpretability, and regularity qualifications. 
In our notation, $\bF$ is the true function that generated the data whereas $\bfun$ is our model for $\bF$; it is hoped that $\bF$ and $\bfun$ share some meaningful properties.
The function $\bfun$ is typically restricted to a given class of models (e.g., linear, polynomial, etc.), so that it may be written as the expansion
\begin{align}\label{Eq:SimpleExpansion}
    \bfun(\bx) \approx \sum_{j=1}^N \bxi_{j}\phi_j(\bx) \quad\Longrightarrow\quad \bfun(\bx) \approx \bXi \bphi(\bx).
\end{align}
Here, $\bphi$ describes the feature library of $N$ candidate terms that may describe the dynamics, and $\bXi$ contains the coefficients that determine which model terms are active and in what proportions.  

Mathematically, the optimisation problem to be solved is
\begin{align}
	\argmin_{\bXi} \left\| \bY - \bXi\bphi(\bX) \right\|_F 
	+ \lambda R(\bXi)%\left\|\bfun \right\|
	\label{Eq:op2}
\end{align}
where $\|\cdot\|_F$ is the Frobenius (Hilbert--Schmidt) norm. 
The first term in \eqref{Eq:op2} corresponds to the error between training samples and our model prediction, whereas the second term  $\lambda R(\bXi)$ is a regularizer. 
For example, in SINDy, the feature library $\bphi$ will typically include linear and nonlinear terms, and the regularizer will involve the number of nonzero elements $\|\bXi\|_0$, which may be relaxed to the $1$-norm $\|\bXi\|_1$. 
In DMD, the features $\bphi$ will simply contain the state $\bx$, $\bXi$ will be the DMD matrix $\bA$, and instead of a regularizer $R(\bXi)$, the minimization is constrained so that the rank of $\bA=\bXi$ is equal to $r$. 
Similarly, in extended DMD, the feature $\bphi$ will include nonlinear functions of the state, and the minimization is modified to $\argmin_{\bXi} \|\bphi(\bY)-\bXi\bphi(\bX)\|_F+\lambda R(\bXi)$, resulting in a $\bXi$ that is a large square matrix evolving the nonlinear feature space forward in time.  

\begin{figure}[t]
\vspace{-.1in}
\begin{center}
	\includegraphics[width=.9\linewidth]{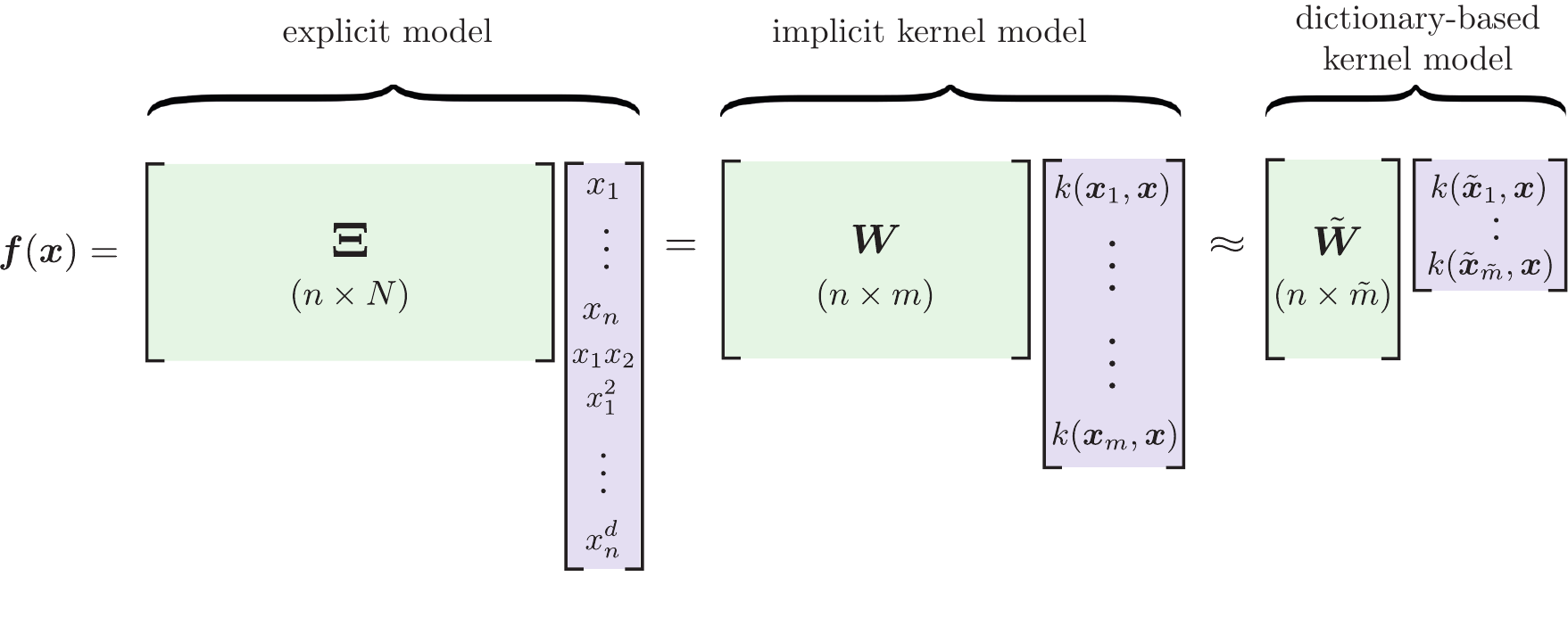}
\end{center}
\vspace{-.45in}
\caption{
Schematic relationships between different models for $N \gg n, m \gg \tilde{m}$.
An explicit model (e.g. SINDy) produces explicit weights that connect $N$ features to $n$ outputs.
A kernel model uses fewer weights but the
relationships between variables are stored implicitly.
The dictionary-based kernel model selects the most active
samples and therefore uses fewer weights still.
}%
	\label{Fig:kernelRegression}
\end{figure}
For even moderate state dimensions $n$ and feature complexity, such as the order of monomials $d$, the feature library of $\bphi$ becomes prohibitively large and the optimization in \eqref{Eq:op2} is intractable.  
This scaling issue is the primary challenge in applying SINDy to high-dimensional systems.  
Instead, it is possible to rewrite the expansion \eqref{Eq:SimpleExpansion} in terms of an appropriate kernel function $k$ as
\begin{align}\label{Eq:KernelExpansion}
    \bfun_i(\bx) \approx \sum_{j=1}^m w_{ij} k(\bx_j,\bx) \quad\Longrightarrow\quad \bfun(\bx) \approx \bW k(\bX,\bx).
\end{align}
In this case, the sum is over the number of snapshots $m$ instead of the number of library elements $p$, dramatically improving the scaling.   
The optimization in \eqref{Eq:op2} now becomes 
\begin{align}
	\argmin_{\bW} \left\| \bY - \bW k(\bX,\bX) \right\|_F 
	+ \lambda R(\bW).%\left\|\bfun \right\|
	\label{Eq:op2kernel}
\end{align}
We will show that it is possible to improve the scaling further by using a kernel defined on a sparse dictionary $\tilde{\bX}$.   
Figure~\ref{Fig:kernelRegression} shows our dictionary-based kernel modeling procedure, where the explicit model on the left is a SINDy model, and the compact model on the right is our kernel model.  
Thus, our kernel learning approach may be viewed as a kernelized SINDy without sparsity promotion. 

Based on the implicit LANDO model, it is possible to efficiently extract the linear component $\bL$ of the dynamics, along with a matrix for the nonlinear forcing:
\begin{align}\label{Eq:LANDODecomposition}
    \bY = \bL\bX + \bN
\end{align}
where here $\bN = \begin{bmatrix}\bN(\bx_1) & \bN(\bx_2) & \cdots & \bN(\bx_m) \end{bmatrix}$ is a nonlinear snapshot matrix, where each column is the nonlinear component of the dynamics at that instant in time.  
Although this is not an explicit expression for the nonlinear dynamics, as in SINDy, knowing the linear model and nonlinear forcing will enable data-driven resolvent analysis~\cite{herrmann2021jfm}, even for strongly nonlinear systems. 
Technically, \eqref{Eq:LANDODecomposition} may be centered at any base point $\bar{\bx}$, resulting in
\begin{align}
    \by_j' = \bL'\bx_j' + \bN(\bx_j')
\end{align}
where $\bx' = \bx - \bar{\bx}$.  We will also show that the linear model $\bL$ may be represented efficiently without being explicitly constructed, as in DMD. 

\begin{figure}[t]
	\centering
	\includegraphics[width = \linewidth]{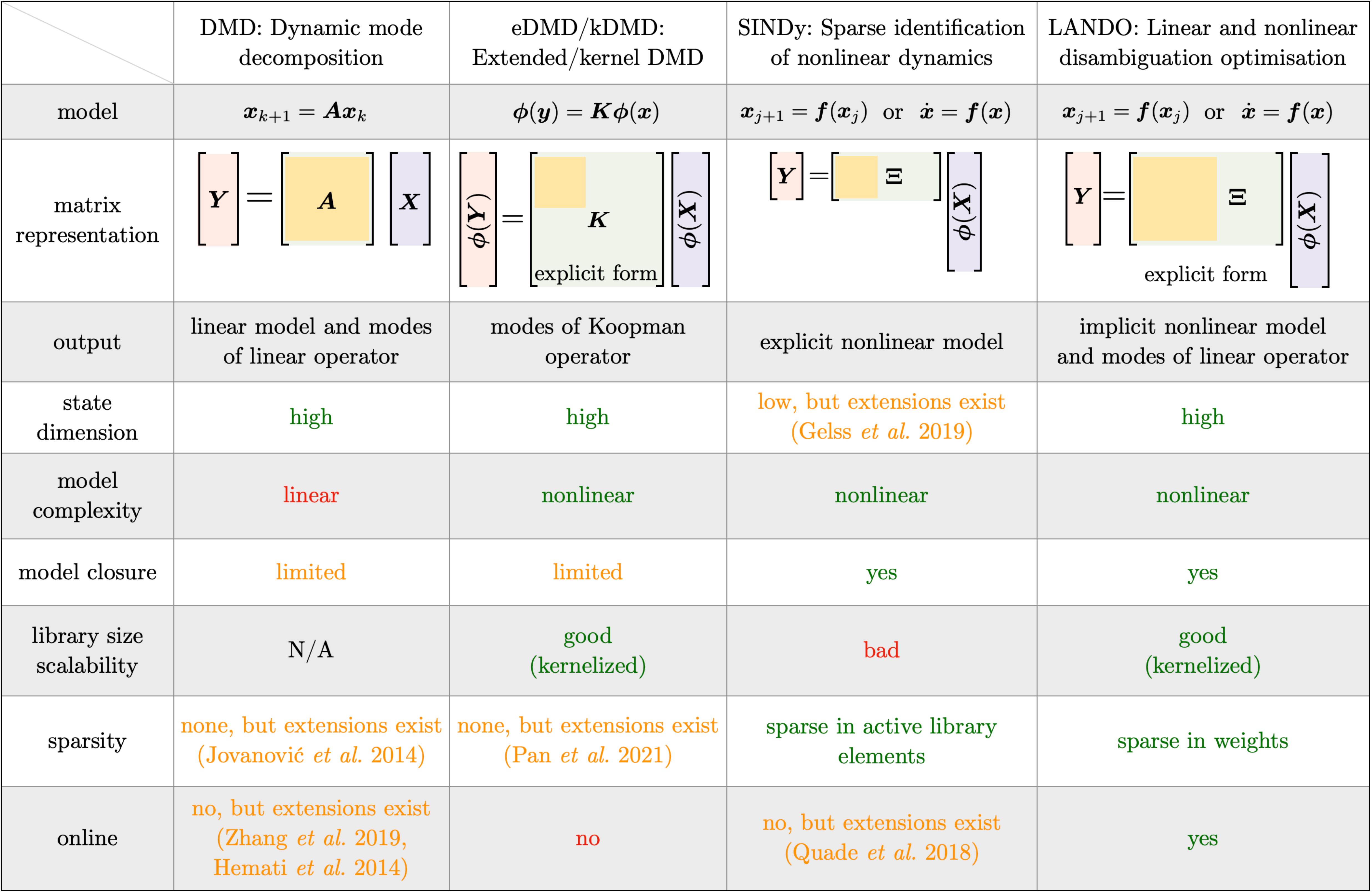}
	\caption{
A comparison of methods for model discovery, including DMD~\cite{Rowley2009,Schmid2010}, extended/kernel DMD~\cite{Williams2015a,Williams2015b}), SINDy~\cite{Brunton2016}, and the proposed LANDO framework.}
	\label{Fig:methodsComparison}
\end{figure}

Figure~\ref{Fig:methodsComparison} compares the proposed LANDO architecture with the leading data-driven model regression techniques it builds upon: DMD, extended/kernel DMD, and SINDy. 
The extended DMD algorithm has already been kernelized~\cite{Williams2015a,Williams2015b}, enabling efficient approximations to the Koopman operator with very large feature spaces.  
Although it is related to the present work, the goal of eDMD/kDMD is to obtain a \emph{square} representation of the dynamics of measurement functions in a Hilbert space or feature space $\bphi(\bx)$, rather than a closed representation of the dynamics in the original state $\bx$.  
In this way, our approach more closely resembles the SINDy procedure, but kernelized to scale to arbitrarily large problems.  
We will also show that even though the representation of the dynamics is implicit, it is possible to extract explicit model structures, such as the linear component and other relevant quantities, from the kernel representation.

In the following subsections, we will outline the DMD, eDMD, and SINDy algorithms and provide an introduction to kernel methods that will be used throughout this work. 
\subsection{Dynamic mode decomposition}
\label{Sec:exactDMD}
The original dynamic mode decomposition (DMD) algorithm of \cite{Schmid2010} was developed as a data-driven method for decomposing high-dimensional snapshot data into a set of coherent spatial modes, along with a low-dimensional model for how these mode amplitudes evolve linearly in time. 
As such, DMD may be viewed as a hybrid algorithm combining principal component analysis (PCA) in space and the discrete-time Fourier transform (DFT) in time~\cite{Chen2012a}.  
DMD has been adopted in a wide range of fields beyond fluid mechanics, including epidemiology~\cite{proctor2015ih}, neuroscience~\cite{brunton2016extracting}, video processing~\cite{grosek2014arxiv,erichson2016jrtp}, robotics~\cite{berger2014ieee,abraham2019ieee,bruder2019proc,mamakoukas2019proc}, finance~\cite{mann2016qf}, power grids~\cite{susuki2011c,susuki2011a}, and plasma physics~\cite{taylor2017arxiv,kaptanoglu2020pop}.
Much of this success stems from the formulation of DMD as a linear regression problem~\cite{Tu2014}, based entirely on measurement data, resulting in several powerful extensions~\cite{Kutz2016}, including for control~\cite{Proctor2016}, sparsity promoting DMD~\cite{Jovanovic2014}, recursive DMD~\cite{Noack2016jfm}, multi-resolution analysis~\cite{Kutz2016siads}, streaming data~\cite{hemati2014pof}, for non-sequential time-series~\cite{Tu2014,Askham2018} and for data that is under-resolved in space~\cite{Brunton2015jcd,Gueniat2015pof} or time~\cite{Tu2014ef}. 

The original algorithm was refined by \cite{Tu2014} who phrased DMD in terms of the 
Moore--Penrose pseudoinverse thereby allowing snapshots that are not equally spaced in time;
this variant is called {\it exact DMD} and will be the main form of DMD used in this paper.
In appendix \ref{Ap:DMDcomparison} we will show that exact DMD may be viewed as a special case of our new method.

As mentioned in the previous section, it is assumed that each $\bx_j$ 
and $\by_j$ are connected by a dynamical system of the form $\by_j = \bF(\bx_j)$.
The aim of DMD is to learn information about $\bF$ by approximating it as a linear operator and
then performing diagnostics on that approximation.
In particular, DMD seeks the linear operator $\bA$ that best maps the sets $\{ \bx_j \}$ and $\{ \by_j \}$ 
into one another:
\begin{align}
\by_j \approx \boldsymbol{A}\bx_j \qquad \qquad \textrm{for } j = 1, \dots, m. 
	\label{Eq:dmd1}
\end{align}
Expressed in terms of the snapshot matrices in \eqref{Eq:SnapshotMatrices}, \eqref{Eq:dmd1} becomes
\begin{align}
{\boldsymbol{Y} \approx \boldsymbol{A}\boldsymbol{X}},
\end{align} 
and the minimum norm solution (without regularisation) is
\begin{equation}
\bA = \argmin_{\bA}\|\bY-\bA\bX\|_F = 
\bY \bX^{\dagger}
\label{Eq:exactDMD}
\end{equation}
where $\dagger$ indicates the Moore--Penrose pseudoinverse \citep{Golub2013}.
If $\bX$ has the singular value decomposition
\begin {equation}
\boldsymbol{X}=\boldsymbol{U\Sigma V}^\ast
\end{equation}
then
\begin{align}
	\bA = \bY \bV \bSigma^\dagger \bU^\ast.
\end{align}
Note that $\bA$ is an $n\times n$ matrix 
so may be extremely large in practical scenarios where $n\gg 1$.
Thus, it is common to use a rank-$r$ approximation for $\bA$,
denoted by $\hat{\bA}$, where $r \ll n$.
To construct $\hat{\bA}$, we construct a rank $r$ approximation for $\bX$ using the truncated singular value decomposition:
$\bX\approx\bU_r\bSigma_r \bV_r^\ast$.
This approximation is optimal according to the Eckart--Young theorem \citep{Eckart1936}.
The matrix $\boldsymbol{A}$ is then projected onto the column space of $\bX_r$ as
\begin{equation}
\hat{\bA} = \bU_r^\ast \bA \bU_r 
= \bU_r^\ast \bY \bV_r \bSigma_r^{-1}.
\end{equation}
Since $\hat{\bA}$ is an $r\times r$ matrix, it is now feasible to compute its eigendecomposition as
\begin{equation}
\hat{\bA} \hat{\bPsi} = \hat{\bPsi} \boldsymbol{\Lambda}.
\end{equation}
It was proved by \cite{Tu2014} that the eigenvectors of the full 
matrix $\bA$ can be approximated from the reduced eigenvectors $\bPsi$ by
\begin{equation}
\bPsi = \bY \bV \bSigma^{-1} \hat{\bPsi}.
\end{equation}
This eigendecomposition has many favourable properties.
Firstly, it is an approximation to the spectrum of the underlying Koopman operator of the system \citep{Rowley2009}.
Secondly, if the snapshots are equally spaced in time and $\by_j = \bx_{j+1}$ then 
the data can be reconstructed in terms of the eigenvectors and eigenvalues as
\begin{align}
\bx_j = \bPsi \bLambda^{j-1} \ba
\end{align}
where the vector $\ba$ contains the mode amplitudes often computed as $\ba = \bPsi^{\dagger}\bx_1$.
The above provides a clear physical interpretation of the modes:
the eigenvectors $\bPsi$ are the spatial modes whereas the eigenvectors 
$\bLambda$ correspond to the temporal component.

\subsubsection{Sparse identification of nonlinear dynamics}
The SINDy algorithm~\cite{Brunton2016} was developed based on the observation that many complex dynamical systems may be expressed as systems of differential equations with only a few terms, so that they are \emph{sparse} in the feature space $\bphi(\bx)$.  
Thus, it is possible to solve for an expansion of the dynamics in \eqref{Eq:SimpleExpansion} with only a few nonzero entries in $\bXi$, corresponding to the active terms in the $\bphi(\bx)$ that are present in the dynamics.  
Solving for the sparse vector of coefficients, and therefore the dynamical system, is achieved through the following optimization
\begin{align}
	\argmin_{\bXi} \left\| \bY - \bXi\bphi(\bX) \right\|_F 
	+ \lambda \|\bXi\|_0.%\left\|\bfun \right\|
	\label{Eq:SINDyOpt}
\end{align}
The $\|\cdot\|_0$ term is not convex, although there are several relaxations that yield accurate sparse models.  
The SINDy algorithm has also been extended to include partially known physics~\cite{Loiseau2018}, such as conservation laws and symmetries, dramatically improving the ability to learn accurate models with less data.  
It is also possible with SINDy to disambiguate the linear and nonlinear model contributions, enabling linear stability analyses, even for strongly nonlinear systems.  
However, the feature library $\bphi(\bx)$ scales poorly with the state dimension $n$, so SINDy is typically only applied to relatively low-dimensional systems.  
A recent tensor extension to SINDy~\cite{Gelss2020Kernel} provides the ability to handle much larger libraries, which is very promising. 
In the present work, we use kernel representations to obtain tractable implicit models that may be queried to extract structure, such as the disambiguated linear terms.   

\subsubsection{Extended DMD}
The extended DMD~\cite{Williams2015a} was developed to improve the approximation of the Koopman operator by augmenting the DMD vector $\bx$ with nonlinear functions of the state, similar to the feature vector $\bphi(\bx)$ above.  
However, instead of modeling $\bx_{k+1}$ as a function of $\bphi(\bx_k)$, as in SINDy, eDMD models the evolution of $\bphi(\bx_{k+1})$, which results in a much larger regression problem. 
\begin{align}
	\argmin_{\bXi} \left\| \bphi(\bY) - \bXi\bphi(\bX) \right\|_F .
	\label{Eq:eDMDOpt}
\end{align}
This approach was then kernelized~\cite{Williams2015b} to make the algorithm computationally tractable.  
\subsection{Kernel methods}
\label{Sec:kernel}
Kernel methods are a class of statistical machine learning algorithms that
perform efficient computations with high-dimensional nonlinear features
\citep{Scholkopf2002,Shawe-Taylor2004}.
Kernel methods have found applications in adaptive filtering~\cite{Liu2008,Liu2010},
nonlinear principal component analysis~\cite{Scholkopf1998,Mika1999},
nonlinear regression~\cite{Engel2004},
classification~\cite{Herbrich2001,Camps-Valls2005},
and support vector machines~\cite{Cristianini2000}. 
The broad success of kernel machines stems from their ability to
efficiently compute inner products in a high-dimensional,
or even infinite-dimensional, nonlinear feature space.
Thus, if a conventional linear algorithm can be phrased exclusively in terms of inner products then it can be
``kernelised'' and adapted for nonlinear problems.
This ``kernel trick'' has been used to great effect in the above applications.

Kernels are continuous functions $k:\mathbb{R}^n \times \mathbb{R}^n \rightarrow \mathbb{R}$, and
a kernel is a Mercer kernel if it is positive definite; i.e. for any collection of vectors	
$\bx^\prime_j \in \mathbb{R}^{n}$, the matrix $\bK$ defined by \mbox{$[\bK ]_{i,j} = k(\bx_i^\prime,\bx_j^\prime)$}
is positive definite.
By Mercer's Theorem \citep{Mercer1909} it follows that there exists a Hilbert space $\H_k$
and a mapping $\bphi:\mathbb{R}^n \rightarrow \H_k$ such that $k(\bx, \bx^\prime) = \left<\bphi(\bx),\bphi(\bx^\prime) \right>$.
In other words, every Mercer kernel can be interpreted as an inner product in the Hilbert space $\H_k$, which
may be of an otherwise inaccessible dimension.
Every element $\bg \in \H_k$ can be expressed as a linear combination
\begin{align}
	\bg(\bx) = \sum_{j=1}^M \balpha_j k(\bx_j^\prime,\bx)
	\label{Eq:genSol}
\end{align}
for some $M \in \mathbb{N}$, $\balpha_i\in \mathbb{R}^n$ and $\bx_j^\prime\in \mathbb{R}^n$. We drop the word 'Mercer in the remainder of the article, and assume that all kernels are Mercer kernels.

An important result in the theory of kernel learning is the \emph{representation theorem}.
First proved by \cite{Kimeldorf1971} and then generalised by \cite{Scholkopf2001}, the representation
theorem provides very general conditions where kernel methods can be used to solve machine learning problems.
For the purposes of the present work, the representation theorem may be stated thus:
for a set of pairs of $m$ training samples, $(\bx_1, \by_1),\, \dots \,, (\bx_m, \by_m)$, the solution to the minimisation problem
\begin{align}
	\argmin_{\bfun \in \H_k}
	\left\|\bY - \bfun(\bX) \right\|_F + \lambda R(\bfun) % \left\|\bfun \right\| 
	\label{Eq:min}
\end{align}
may be expressed as
\begin{align}
	\bfun(\bx) = \sum_{j=1}^m \bw_j k(\bx_j, \bx)
	\label{Eq:rep}
\end{align}
for vectors $\bw_j \in \mathbb{R}^n$.
One important consequence of the representation theorem is that the solution to the optimisation problem \eqref{Eq:min} can be expressed as a linear combination of kernel functions whose first arguments are the training data.
Contrast this with the general representation of members of $\H_k$ in \eqref{Eq:genSol} where the parameters $\bx_j^\prime$ are not known.
The representation theorem allows us to avoid an exhaustive search for the optimal parameters, thereby reducing the problem to a (linear) search for the weights $\bw_j$.
In the above, $\lambda>0$ is a regularisation parameter and the regulariser on $\bfun$ is to be interpreted as the norm associated with $k$:
\begin{align}
R(\bfun) = \| \bfun \|^2
=	\left\| \sum_{j=1}^m \bw_j k(\bx_j,\cdot) \right\|^2
	= \sum_{i=1}^m \sum_{j=1}^m \bw_j^\ast \bw_i k(\bx_j,\bx_i).
\end{align}
\subsubsection{An illustrative example}
\label{Sec:kernelExample}
The discussion of kernels has thus far been rather abstract;
we now make the theory concrete by illustrating
an application of the usefulness of kernel methods.
This simple example is often used
in kernel tutorials \cite{Scholkopf2002,Williams2015b}.

Consider a three-dimensional state, $\bx \in \mathbb{R}^3$, upon which we want to perform some machine learning task such as regression or classification.
Suppose that we know -- from either physical intuition, empirical data, or experience --
that the system is governed by pairwise quadratic interactions between the state variables. 
Thus, our machine learning model should operate in the nonlinear feature space defined by
\begin{align}
	\bphi(\bx) = \left[x_1 x_2\quad x_1 x_3 \quad x_2 x_3 \quad x_1^2 \quad x_2^2 \quad  x_3^2 \right]^T
	\in \mathbb{R}^{6}.
\end{align}
Almost every machine learning algorithm uses inner products to measure correlations between samples.
Computing inner products in a feature space of dimension $N$  costs $2N - 1$ operations:
$N$ products and $N-1$ summations.
Thus, in this example, computing inner products in the nonlinear feature space would usually require $11$ operations.
However, we still need to form the feature vector $\bphi(\bx)$, which costs a further 6 products, raising the total count to 17 operations.

Equivalently, we could build our model in the slightly rescaled feature space:
\begin{align}
	\boldsymbol{\varphi}(\bx) = 
 \left[\sqrt{2}x_1 x_2\quad \sqrt{2}x_1 x_3 \quad \sqrt{2} x_2 x_3 
	\quad x_1^2 \quad x_2^2 \quad  x_3^2 \right]^T.
\end{align}
Now note that inner products in this feature space may be expressed as
\begin{align}
	\left<\boldsymbol{\varphi}(\bx),\boldsymbol{\varphi}(\bx^\prime)\right>
	&= 2 x_1 x_1^\prime x_2 x_2^\prime
+2 x_1 x_1^\prime x_3 x_3^\prime 
+2 x_2 x_2^\prime x_3 x_3^\prime 
+x_1^2 (x_1^\prime)^2
+x_2^2 (x_2^\prime)^2
+x_3^2 (x_3^\prime)^2\notag\\
	&= \left( x_1 x_1^\prime + x_2 x_2^\prime + x_3 x_3^\prime\right)^2\notag\\
	&= \left(\left< \bx, \bx^\prime\right>\right)^2.
	\label{Eq:kern1}
\end{align}
Thus, we can compute the inner product $\left<\boldsymbol{\varphi}(\bx),\boldsymbol{\varphi}(\bx^\prime)\right>$
in merely $6$ operations by computing the inner product $\left< \bx, \bx^\prime\right>$ and then squaring the result.
In other words, computing the inner product amounted to evaluating the kernel $k(\bu,\bv) = (\bu^T \bv)^2$.
Moreover, while computing the inner product with the kernel, we never explicitly formed the feature space,
and therefore did not need to store $\boldsymbol{\varphi}$ in memory.
In summary, if we use expression \eqref{Eq:kern1} 
then the cost of computing inner products falls from 17 operations to 6 operations.

This may seem a modest saving but the cost of computations
in feature space explodes as the state dimension or degree of nonlinearity increase.
For a state of dimension $n$, the number of degree $d$ monomial features is%
\footnote{
This can be thought of as the number of ways of distributing $d$ unlabelled balls 
into $n$ labelled urns.}
\begin{align}
	N = \binom{n + d - 1}{d} = \frac{(n + d -1)!}{d!\, (n-1)!}.
\end{align}
Thus, explicitly forming vectors in this feature space is extremely expensive,
as is computing inner products.
For example, for an $n$-dimensional state, the number of possible quadratic interactions between states
is $n(n-1)/2$. 
This scaling of the feature vector is the prime limitation of SINDy. 

Instead of explicitly forming this vast feature space we instead work with suitably chosen kernels.
The feature space of degree $d$ monomials can be represented using the \emph{polynomial kernel}
\begin{align}
	k(\bu,\bv) = (\bu^T \bv)^d.
	\label{Eq:monKernel}
\end{align}
Thus, using the kernel \eqref{Eq:monKernel} to compute inner products reduces the operation count from
$2 N-1$ to $2n$, which is significant when the state space is large and the nonlinearity is quadratic or higher.
\section{Learning kernel models with sparse dictionaries}
\label{Sec:method}
We now develop the main learning method presented by this paper.
The procedure is based on the kernel recursive least squares (KRLS) algorithm of~\cite{Engel2004} but is more stable and allows further interpretation and analysis of the learned model. We specifically tailor this approach to learn dynamical systems in a robust and interpretable framework. 
Recall that we are solving the optimization problem defined in \eqref{Eq:op2} for a nonlinear function $\bfun$ that approximates the dynamics.
By the representer theorem, we may express the dynamical system approximation $\bfun$ from \eqref{Eq:SimpleExpansion} in the kernelized form \eqref{Eq:rep} as:
\begin{align}
    \bfun(\bx) \approx \sum_{j=1}^p \bxi_{j}\phi_j(\bx) = \sum_{j=1}^m\bw_j k(\bx_j,\bx).
\end{align}
Arranging the column vectors $\bw_j$ into a matrix $\bW$ 
allows us to write $\bfun(\bx) = \bW \, k(\bX, \bx)$ so the optimisation problem is
\begin{align}
	\argmin_{\bW}
	\left\|\bY - \bW \, k(\bX, \bX) \right\|_F + \lambda R(\bfun).%\left\|\bfun \right\|.
	\label{Eq:problem}
\end{align}
Theoretically, a solution to \eqref{Eq:problem}, in the absence of regularisation, is provided by the Moore--Penrose pseudoinverse:
\begin{align}
	\bW = \bY  k(\bX, \bX)^\dagger.
	\label{Eq:fullSol}
\end{align}
As noted by~\cite{Engel2004}, there are three practical problems with the above solution:
\begin{itemize}
\item \myuline{Numerical conditioning}: even though the kernel matrix may formally have full rank, it will usually have a very large condition number
since the samples can be almost linearly dependent in the feature space.
When the condition number is large, the condition number of the pseudoinverse will also be large and $\bW$ will amplify noise by a corresponding amount.
\item \myuline{Overfitting}:
The weight matrix $\bW$ has $mn$ entries, which is equal to the number of constraining equations in \eqref{Eq:problem}.
Thus, there is a risk of overfitting, which can limit the generalisability of the model and make it sensitive to noise.
\item \myuline{Computational complexity}:
	In nonlinear system identification we usually need a large number of samples to adequately
	learn the system. When there are $m\gg1$ samples,
	constructing the pseudoinverse $k(\bX,\bX)^\dagger$ requires $O(m^3)$ operations to construct and $O(m^2)$ space in memory, which can become prohibitively expensive.
	Additionally, evaluating the model $\bfun$ for prediction or reconstruction requires multiplying
	the $m\times n$ weight matrix by the $m$-vector of kernel evaluations, which will also become expensive for large sample sets.
\end{itemize}
To address these issues, Engel \emph{et al.}~\cite{Engel2004} proposed an online form of dimensionality reduction
that iteratively constructs a dictionary of samples that capture the salient features of the underlying dynamics.
The key idea is that the model $\bfun$ defined in \eqref{Eq:rep} can be approximated by
\begin{align}
	\bfun(\bx) \approx \tilde{\bW}\,\, k(\tilde{\bX},\bx)
	\label{Eq:fApprox}
\end{align}
for a suitable choice of $\tilde{\bX}$ known as the \emph{dictionary}; in this paper the tilde symbol indicates that a quantity is connected to the dictionary.
Then, the optimisation \eqref{Eq:problem} may be approximated as 
\begin{align}
	\argmin_{\tilde{\bW}}
	\left\|\bY - \tilde{\bW} \, k(\tilde{\bX}, \bX) \right\|_F + \lambda R(\bfun).%\left\|\bfun \right\|.
	\label{Eq:problemDictionary}
\end{align}
The dictionary is constructed by considering each sample and determining whether they should be included in the dictionary.
Membership of a sample in the dictionary is decided by checking if the sample can be approximated in the feature space using the current dictionary.
This scheme is called the `almost linearly dependent' (ALD) test: if a sample is approximately linearly dependent on the current dictionary then it is not added, otherwise the dictionary must be updated with the current sample. 
Thus, the dictionary is a sparse\footnote{We use the term `sparse' carefully here. The dictionary is actually a dense matrix but consists of a small number of the total samples. This is the terminology used in the original work of~\cite{Engel2004}.}
subset of samples that spans the largest subspace in the data.
Usually, the size of the dictionary is much smaller than the number of samples.

The dictionary learning procedure searches the high-dimensional feature space for a low-dimensional subspace where most of the dynamics take place.
This approach is similar to kernel principal component analysis (KPCA,~\cite{Scholkopf1998}), though we argue that ALD dictionary learning is more physically interpretable.
KPCA conflates the feature space representations of samples, and the result usually has no interpretation in the original physical space.
For example, if the feature space is $\bphi(\bx) = [x_1\;\; x_2 \;\; x_1 x_2]^T$ then certain datasets could produce a principal component of $\hat{\bphi} = [1\;\; 1\;\; 0]^T$.
However, such a vector is unrealisable in the original physical space because if the $x_1x_2$ component is zero then at least one of $x_1$ and $x_2$ must also be zero.
Thus, there is no clear interpretation of the principal components in the original state space; indeed, as we have demonstrated, the principal components may be mathematically inconsistent.
The problem of identifying the state space vector that is closest to a given principal component in the feature space is known as the \emph{pre-image problem}~\cite{Mika1999,Kwok2004} and is usually solved via an iterative optimisation procedure.
It was shown in \cite{Engel2004} that ALD dictionary learning may be viewed as an approximate form of KPCA.
Additionally, the dictionary has a clear physical interpretation since every member it contains is simply the state vector system at a specific time. 
Thus, ALD dictionary learning may be preferable to KPCA when studying physically motivated problems.

\subsection{Sparse dictionary learning}
\label{Sec:dictionary}
The dictionary at time $t$ is defined as a collection of 
$\tilde{m}_t$ vectors, $\mathcal{D}_t = \{ \tilde{\bx}_j \,|\, j = 1, \cdots, \tilde{m}_t \}$, and is initialised with $\mathcal{D}_1 = \{ \bx_1\}$.
We write
\begin{align}
	{{\bX}_t=  \begin{bmatrix} |  & | & & |
		\\ {\bx}_1 & {\bx}_2 &  \cdots & 
{\bx}_{t} \\ | & | &  & |\end{bmatrix}},
\qquad
	{{\bY}_t=  \begin{bmatrix} | & | &  & |
		\\ {\by}_1 & {\by}_2 & \cdots &
{\by}_{t} \\ | & | & & |\end{bmatrix}}
\end{align}
to represent data matrices including all samples up to snapshot $t$.
We may also represent the dictionary in terms of data matrices as
\begin{align}
	{\tilde{\bX}_t=  \begin{bmatrix} | & | & & |
		\\ \tilde{\bx}_1 & \tilde{\bx}_2 &  \cdots & 
\tilde{\bx}_{\tilde{m}_t} \\ | & | &  & |\end{bmatrix}},
\label{Eq:dictionaryDataMat}
\end{align}%
and in feature space as
\begin{align}
	{\tilde{\bPhi}_t=  \begin{bmatrix} | & | &  & |
		\\ \bphi\left(\tilde{\bx}_1\right) & \bphi\left(\tilde{\bx}_2\right) & \cdots & 
\bphi\left(\tilde{\bx}_{\tilde{m}_t}\right) \\ | & | &  & |\end{bmatrix}}.
\end{align}
When a new element is introduced, we determine how much new information it could add to our model.
In other words, how well can the new element be approximated using the members of the current dictionary.
The degree to which the current sample can be well-represented by the dictionary
in feature space is quantified by
\begin{align}
	\delta_t = \min_{\bpi_t} \left\| \bphi(\bx_t) - \tilde{\bPhi}_{t-1}\bpi_t \right\|_2^2.
\end{align}
The number $\delta_t$ represents the minimum distance between the current sample and the span of the current dictionary and $\bpi_t$ specifies the linear combination of dictionary elements that minimises this distance.
Having calculated $\delta_t$, detailed below, we compare it to a user-defined sparsification threshold $\nu$.
If $\delta_t\leq\nu$ then the new sample $\bx_t$ can be approximated in the feature space using linear combinations of members of the dictionary.
Thus, the new sample is ALD on the dictionary elements in the implicit feature space.
If $\delta_t>\nu$ then the new sample cannot be well-approximated by the current dictionary. % with error less than $\nu$.
Thus, the new sample contributes meaningful information that was not already present in the dictionary and the dictionary should be updated with the current sample.

By expanding the norm and using properties of kernels, we can show that
\begin{align}
	\delta_t =  k_{tt} - \tilde{\bk}_{t-1}^\ast \bpi_t
	\label{Eq:delta}
\end{align}
where the minimiser is
\begin{align}
\bpi_t = \tilde{\bK}_{t-1}^{-1} \tilde{\bk}_{t-1}
\label{Eq:smallPiUpdate}
\end{align}
and
\begin{align}
	k_{tt} = k(\bx_t, \bx_t),
	\qquad \qquad
	\tilde{\bk}_{t-1} = k(\tilde{\bX}_{t-1},\bx_t) \in \mathbb{R}^{\tilde{m}_{t-1}}.
	\label{Eq:kTilde}
\end{align}
The kernel matrix $\tilde{\bK}_{t-1}^{-1}$ and its inverse should be updated whenever an element is added to the dictionary.
The update equations are respectively
\begin{align}
	\tilde{\bK}_{t} = 
	\begin{bmatrix}
		\tilde{\bK}_{t-1} & \tilde{\bk}_{t-1}\\[2ex]
		\tilde{\bk}_{t-1}^\ast & k_{tt}
	\end{bmatrix},
\qquad
\tilde{\bK}_{t}^{-1} = 
	\begin{bmatrix}
		\tilde{\bK}_{t-1}^{-1} + \bpi_t \bpi_t^\ast/\delta_t 
		& -\bpi_t/\delta_t\\[2ex]
		-\bpi_t^\ast/\delta_t& 1/\delta_t
	\end{bmatrix}.
	\label{Eq:Kupd2}
\end{align}
The above expression for $\tilde{\bK}_t^{-1}$ is mathematically correct but numerically unstable.
This is a source of rounding errors when working with the polynomial kernels that we use in practice.
This instability arises through the large condition number of the matrix $\tilde{\bK}_t$.
This issue is typical of kernel methods, which are often plagued with problems of numerical stability due to the large condition numbers associated with kernel matrices.
This seems to not be an issue for the Gaussian kernels that were used in the original KRLS formulation of \cite{Engel2004}, but it becomes important when working with the polynomial kernels that arise in physical applications.
To circumvent these issues, we avoid constructing the ill-conditioned matrices
$\tilde{\bK}_{t}$ and $\tilde{\bK}_{t}^{-1}$ explicitly.
In particular, as $\tilde{\bK}_{t}$ is positive definite it admits a unique Cholesky decomposition
$\tilde{\bK}_t = \chol_t \chol_t^\ast$ 
where $\chol_t$ is a lower-triangular $\tilde{m}_t \times \tilde{m}_t$ matrix \citep{Golub2013}.
% Then, multiplication by the inverse matrix $\tilde{\bK}_t^{-1}$ can be interpreted and implemented
% as a pair of backsubstitutions with $\chol_t$.
Instead of updating $\tilde{\bK}_{t}$ according to \eqref{Eq:Kupd2}, we instead update and store its
Cholesky factor.
The Cholesky factor is initialised as \mbox{$\chol_1 = \sqrt{k_{11}}$} and the update rule is
\begin{align}
	\chol_t = \begin{bmatrix}
		\chol_{t-1} &  \boldsymbol{0} \\[2ex]
		\bs_t^\ast & c_t
	\end{bmatrix}
	\label{Eq:cholUpdate}
\end{align}
where $\bs_t = \chol_{t-1}^{-1} \tilde{\bk}_t$ can be formed in $\mathcal{O}(\tilde{m}_t^2)$ operations by backsubstitution and  $c_t=\sqrt{k_{tt} - \bs_{t}^\ast \bs_t}$. 
Rounding errors can still accumulate and produce an imaginary value for $c_t$, so in practice one can use $c_t = \max \left(0,\sqrt{k_{tt} -\bs_{t}^\ast\bs_t} \right)$.
In summary, multiplication by the inverse $\tilde{\bK}^{-1}_t$ should be interpreted and implemented as solving a linear system with two back substitutions of $\chol_t$.
Thus, we can compute the distance of a sample from the dictionary \eqref{Eq:delta} without ever forming the ill-conditioned matrices $\tilde{\bK}_{t}$ and $\tilde{\bK}_{t}^{-1}$.
The full dictionary can be learned in $\mathcal{O}(m \tilde{m}^2 + n m \tilde{m})$ time.

\begin{figure*}[tpb]
	\vspace{-.15in}
	\centering
	\begin{subfigure}[b]{.45\textwidth}
% 		\hspace{.04cm}
	 \includegraphics[height = 6.cm]{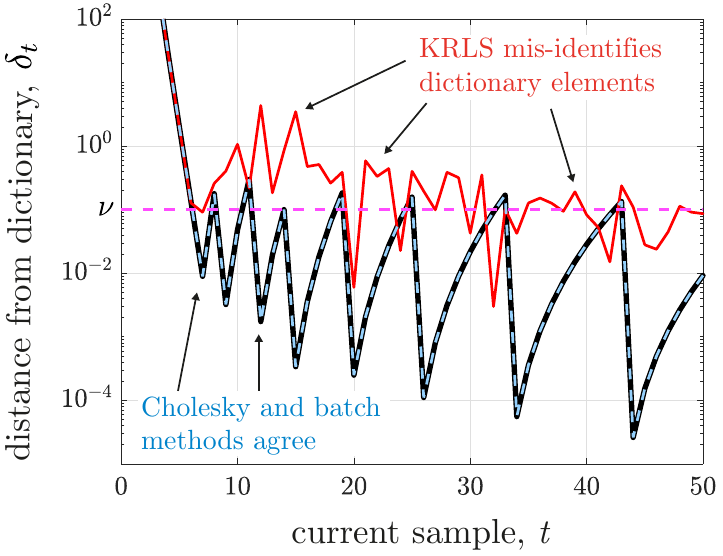}
	\caption{}
\label{Fig:stabilityA}
	\end{subfigure}
	\hfill
	\begin{subfigure}[b]{.45\textwidth}
		%\hspace{-.7cm}
	 \includegraphics[height = 5.95cm]{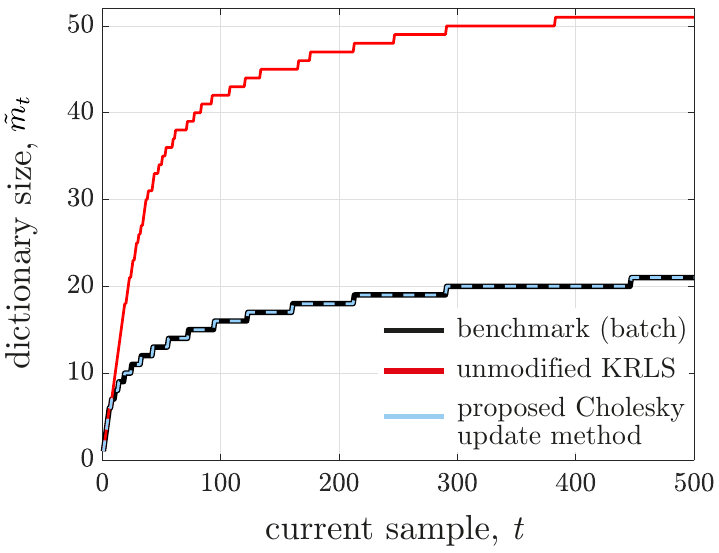}
	\caption{}
\label{Fig:stabilityB}
	\end{subfigure}
	\vspace{-.125in}
	\caption{Comparing the ALD dictionaries computed by the original KRLS algorithm, the Cholesky updating variant and a batch offline algorithm when applied to a solution of the viscous Burgers' equation.
	The kernel here is quadratic and the sparsity parameter is $\nu = 0.1$.
	Figure (a) shows the computed distance of each sample from the span of the current dictionary, which determines whether the current sample should be added to the dictionary.
	Figure (b) plots the growth the dictionary
	as more samples are considered.
	The original KRLS algorithm mis-idenfities dictionary elements
	and the corresponding dictionary is larger than necessary.}%
	\label{Fig:stability}
		\vspace{-.1in}
\end{figure*}

Updated Cholesky factors significantly improve dictionary learning.
Figure \ref{Fig:stability} illustrates the improvement by comparing three methods of dictionary learning.
We evaluate the efficacy of each method by their accuracy in computing $\delta_t$ for each sample, which represent the algorithm's estimate of the distance of sample $t$ from the current dictionary.
Recall that $\delta_t$ determines whether the current sample should be included in the dictionary
so accurate computation of $\delta_t$ is essential.
The first method is the original KRLS formulation, which uses \eqref{Eq:Kupd2} to compute $\delta_t$ with \eqref{Eq:smallPiUpdate}.
The second method is the Cholesky updating formulation presented here, which uses \eqref{Eq:cholUpdate} to compute $\delta_t$ as opposed to constructing $\tilde{\bK}_t$ or $\tilde{\bK}_t^{-1}$ explicitly.
The third method is a batch method that computes $\tilde{\bK}_{t-1}^{-1}$ from scratch at each iteration and doesn't use the estimates of $\tilde{\bK}_t^{-1}$ or $\chol_t$ at the previous iteration.
We take the batch method to be the ground truth, although there will still be some numerical instability associated with the large condition number of $\tilde{\bK}_t$. Although accurate, the batch method is also prohibitively expensive at large scales, with each iteration costing $\mathcal{O}(\tilde{m}^3)$ as opposed to the updating methods which cost merely $\mathcal{O}(\tilde{m}^2)$.
The data here are chronologically ordered samples from a simulation of the viscous Burgers' equation (see Sec.~\ref{Sec:Burgers}), and we use a quadratic kernel with sparsity threshold $\nu = 0.1$.
Figure \ref{Fig:stabilityA} indicates that the first seven samples are all added to the dictionary. 
After this transient period, most new samples are well-represented by the current dictionary and are therefore excluded.
Occasionally, the data drift sufficiently far away from the dictionary that a new sample must be included.
This is illustrated by the spikes appearing in the batch method and the Cholesky updating method in figure \ref{Fig:stabilityA}.
Physically, this indicates that the solution of the PDE has departed from what can be adequately described by the dictionary of previous samples.
The results indicate that the batch method and Cholesky update method select identical dictionaries, whereas the KRLS dictionary learning algorithm mis-identifies a large number of dictionary elements. 
Moreover, figure \ref{Fig:stabilityB} shows that the KRLS dictionary is more than twice the size of the correct dictionary.
In summary, figure \ref{Fig:stability} indicates that the Cholesky factor method significantly improves the accuracy of the learned dictionaries.

In the feature space, we may express all the samples up to time $t$ as
\begin{align}
	\bPhi_t = \tilde{\bPhi}_t \bPi_t + \bPhi^{res}_t
	\label{Eq:dictApprox}
\end{align}
where %
\begin{align}
	{\bPi}_t=  \begin{bmatrix} | & | &  & |
		\\ \bpi_1 & \bpi_2 & \cdots & 
\bpi_{t}\\ | & | &  & |\end{bmatrix} \in \mathbb{R}^{\tilde{m}_t \times t}
\end{align}
maps the $\tilde{m}_t$ dictionary elements into the
$t$ feature vectors with small residual error $\bPhi^{res}_t$.
By causality, the lower triangular elements of $\bPi_t$ are zero.
Only the online version of the algorithm (see appendix \ref{Ap:online}) uses $\bPi_t$ explicitly,
and only requires $\bpi_t$ at time $t$. 
Thus, $\bpi_t$ can be overwritten at each iteration to save memory.
Pseudocode for the procedure may be found in algorithm \ref{Alg:dictionary}.

{Alternatives to this proposed dictionary learning procedure include randomised methods \cite{Williams,Rahimi2007} and the recent method in \cite{Gelss2020Kernel}.}

\begin{algorithm}[t]
	\caption{Sparse ALD dictionary learning with Cholesky updates \\
		The operation count for each step is included on the right
	}
\label{Alg:dictionary}
\textbf{Inputs}:
	      data matrix $\bX$,
	      kernel $k$,
	      sparsification tolerance $\nu$\\
\textbf{Output}:
the sparse dictionary $\tilde{\bX}$
  \begin{algorithmic}
  \For{$t = 1 \to m$}
  \State Select new sample $\bx_t$
  \State Compute $\tilde{\bk}_{t-1}$ with \eqref{Eq:kTilde}
  \Comment{$\mathcal{O}(\textcolor{col2}{n\tilde{m}_t})$}
  \State Compute $\bpi_t$ with backsubstitution \eqref{Eq:smallPiUpdate}
  \Comment{$\mathcal{O}(\textcolor{col1}{\tilde{m}_t^2})$}
  \State Compute $\delta_t$ using \eqref{Eq:delta}
  \Comment{$\mathcal{O}(\textcolor{col0}{\tilde{m}_t})$}
  \If{$\delta_t \leq \nu$ (almost linearly dependent)}
  \State Maintain the dictionary: $\mathcal{D}_t = \mathcal{D}_{t-1}$
  \Comment{$\mathcal{O}(\textcolor{col1}{\tilde{m}_t^2})$}
  \ElsIf{$\delta_t > \nu$ (not almost linearly dependent)}
  \State Update the dictionary: $\mathcal{D}_t = \mathcal{D}_{t-1} \cup \{\bx_t\}$
  \State Update the Cholesky factor $\chol_t$ using \eqref{Eq:cholUpdate}
  \Comment{$\mathcal{O}(\textcolor{col1}{\tilde{m}_t^2})$}
\EndIf
  \EndFor
  \end{algorithmic}
\end{algorithm}

\subsection{Batch regression learning}
Once the dictionary has been learned, the optimisation becomes the tractable problem defined in \eqref{Eq:problemDictionary}. There are many methods available to find the weights $\tilde{\bW}$ from \eqref{Eq:problemDictionary}; in the absense of further regularisation on $\tilde{\bW}$, we use the Moore--Penrose pseudoinverse:
\begin{align}
	\tilde{\bW} = \bY \,\,  k(\tilde{\bX}, \bX)^\dagger.
	\label{Eq:kernSol}
\end{align}
The computation of the pseudoinverse is far cheaper than the full solution \eqref{Eq:fullSol} and avoids the issues
described earlier.
Thus, we are left with two quantities that together define the nonlinear model in \eqref{Eq:fApprox}: 
the final dictionary matrix $\tilde{\bX}$ with $\tilde{m}$ columns and
the final set of weights $\tilde{\bW}$.

The model may also be learned in a purely online fashion (see appendix \ref{Ap:online}), which is useful when working with streaming data.
The algorithm is also applicable to situations where the system is forced by an exogenous control variable: details are provided in appendix \ref{Ap:control}.
\section{Extracting and enforcing physical structure with kernel machines}\label{Sec:ExtractingStructure}
Having calculated the kernel weights $\tilde{\bW}$, we may now construct our model  ${\bfun(\bx) = \tilde{\bW} \, k(\tilde{\bX},\bx)}$ from \eqref{Eq:fApprox}.
This kernel model is \emph{implicit}:
without further analysis we cannot interpret the model and understand the physical relationships that the model has learned.
In this section we present techniques that extract physically interpretable structures from the kernel model $\bfun$.

\subsection{Extracting structure from kernel machines: the linear operator}
\label{Sec:disambiguate}
One means of providing insight and interpretability is to analyze the linear component of $\bfun$ relative to some state.
In particular, suppose we consider perturbations (not necessarily of small amplitude)
about a \emph{base state} $\overline{\bx}$ which may correspond to the mean of the data, 
an equilibrium solution, or simply the zero vector.
We define the perturbations about the base state as $\bx^\prime$ so that
$\bx = \overline{\bx} + \bx^\prime$.
A typical approach is to seek a representation of our model of the form
\begin{align}
	\bfun(\bx) =  \bc + \bL \bx^\prime + \bN(\bx^\prime)
	\label{Eq:linExt}
\end{align}
where $\bL$ is a linear transformation, $\bc$ is a constant, and
and $\bN$ is a nonlinear operator such that
\begin{align}
	\lim_{\|\bx^\prime \|_2 \rightarrow \boldsymbol{0}} 
	\frac{\left\| \bN(\bx^\prime) \right\|_2}{\left\| \bx^\prime \right\|_2}
	={0}.
	\label{Eq:nonlinCond}
\end{align}
In words, condition \eqref{Eq:nonlinCond} restricts $\bN$ so that it is purely nonlinear {with respect to the base state $\overline{\bx}$.}
If $\bL$ and $\bc$ are known, then rearranging \eqref{Eq:linExt} obtains the nonlinear fluctuations as {$\bN(\bx^\prime) = \bfun(\bx) - \bL(\bx^\prime) - \bc$}.

Numerically computing the linear component of a high-dimensional nonlinear operator can be computationally expensive.
For example, neural networks use stochastic gradient descent to estimate the local slopes of high-dimensional functions for optimization. 
By virtue of our use of kernels, we can extract the linear component analytically.
We will illustrate this first with a simple kernel, and then provide the formulas
for more general kernels.

First, suppose that the implicit feature space is of the form
\begin{align}
	\bphi(\bx) = 
	\begin{bmatrix}
a\\
\bD \bx\\
\bomega(\bx)
	\end{bmatrix}
\end{align}
where $a$ is a constant (possibly zero), $\bD$ is a diagonal transformation that scales the data, 
and $\bomega$ is a nonlinear function that satisfies \eqref{Eq:nonlinCond}.
Such feature spaces arise when using polynomial kernels \eqref{Eq:polyKernel}.
An alternative but equivalent representation of the model \eqref{Eq:fApprox} is 
\begin{align}
	\bfun(\bx) = \tilde{\bW} \,\, \tilde{\bPhi}^\ast \,\, \bphi(\bx).
	\label{Eq:kernelExp}
\end{align}
Thus, $\bfun$ may be expressed as
\begin{align}
	\bfun(\bx) = \tilde{\bW}
\begin{bmatrix}
	a \boldsymbol{1} & \left(\bD \tilde{\bX}\right)^\ast & \bomega(\tilde{\bX})^\ast
\end{bmatrix}
\begin{bmatrix}
	a \\ \bD \bx \\ \bomega(\bx)
\end{bmatrix}
= \tilde{\bW} \left( a^2 \boldsymbol{1} + \left( \bD \tilde{\bX} \right)^\ast \bD \bx + \bomega(\tilde{\bX})^\ast \bomega(\bx) \right)
\end{align}
where $\boldsymbol{1}$ is column vector of $\tilde{m}$ ones.
Simply reading off the term proportional to $\bx$ gives $\bL$ and the constant
term gives $\bc$ as
\begin{align}
	\bL = \tilde{\bW} \tilde{\bX}^\ast \bD^2, \qquad
	\bc = a^2 \tilde{\bW} \boldsymbol{1}.
\label{Eq:linopKRLS}
\end{align}
The constant $\bc$ is the sum of the rows of $\tilde{\bW}$ multiplied by $a^2$.
 For example, in the case of the quadratic kernel 
 $k(\bx,\by) = (1+ \bx^\ast \by)^2$, the feature space is
\begin{align}
	\bphi(\bx) = 
	  \begin{bmatrix} 1 &
		\sqrt{2} x_1& \sqrt{2}x_2 &\dots &\sqrt{2} x_{n}&
		x_{1}^2& \sqrt{2} x_{1} x_2 & 
		\sqrt{2} x_1 x_3& \dots &
		x_2^2& \sqrt{2} x_2 x_3 &\dots & x_n^2
\end{bmatrix}^\ast.
\end{align}
so
\begin{align}
	a = 1, \qquad \bD = \sqrt{2} \bI,
			\qquad \bomega(\bx) = 
			\begin{bmatrix}
		x_{1}^2& \sqrt{2} x_{1} x_2 & 
		\sqrt{2} x_1 x_3& \dots &
		x_2^2& \sqrt{2} x_2 x_3 &\dots & x_n^2
			\end{bmatrix}^\ast.
\end{align}
The above analysis is valid for a specific type of kernel and the base state $\overline{\bx} = \boldsymbol{0}$.
Nevertheless, the underlying idea can be generalised to arbitrary kernels and base states $\overline{\bx}$ by considering the 
Taylor expansion of $\bfun$ about $\overline{\bx}$:
\begin{align}
	\bfun(\bx) = \bfun(\overline{\bx}) +
	\left. \nabla \bfun(\bx) \right|_{\bx = \overline{\bx}}\bx^\prime
		+ \textrm{higher-order terms}.
\end{align}
Thus, $\bfun(\bx)$ can be expressed in the form \eqref{Eq:linExt} where
\begin{align*}
	\bc = \bfun(\overline{\bx}), \qquad \bL = \left. \nabla \bfun(\bx) \right|_{\bx = \overline{\bx}},
\end{align*}
\begin{align*}
	\bN(\bx^\prime) &= \bfun(\bx) - \bfun(\overline{\bx}) -
\left. \nabla \bfun(\bx) \right|_{\bx = \overline{\bx}}\bx^\prime.
\end{align*}
Accordingly, to compute $\bc$, $\bL$, and $\bN$ we  need only compute $\bfun$ and
$\nabla \bfun$.
Since our model consists of linear combinations of kernels \eqref{Eq:fApprox}, the gradient is simply
\begin{align}
	\nabla \bfun(\bx) = \tilde{\bW} \,\, \nabla k(\tilde{\bX},\bx).
\end{align}
The gradient $\nabla k$ can usually be computed analytically in a straightforward manner.
For example, for polynomial kernels \eqref{Eq:polyKernel} we have
\begin{align}
	\nabla k (\tilde{\bX},\bx) = \tilde{\bX}^\ast d \left(\textrm{diag}[c + \tilde{\bX}^\ast \bx]\right)^{d-1}.
	\label{Eq:polyLin}
\end{align}
The special case in \eqref{Eq:linopKRLS} is recovered
by taking $d = 2$, $c=1$, and $\overline{\bx}=\boldsymbol{0}$.
For Gaussian kernels \eqref{Eq:gaussKernel}, the gradient is
\begin{align}
	\nabla k (\tilde{\bX},\bx) 
	=\frac{-1}{\sigma^2}
\tilde{\bX}^\ast
\textrm{diag}\left[ 
\|\tilde{\bx}_j - \bx\|_2 \,\,
\exp\left(
	-\|\tilde{\bx}_j - \bx\|_2^2/(2 \sigma^2)
\right)
	\right].
\label{Eq:gaussLin}
\end{align}
Similar expressions can be derived for any kernel function or any combination of kernels.

Note that the gradients \eqref{Eq:linopKRLS}, \eqref{Eq:polyLin}, and \eqref{Eq:gaussLin} all take the form $\nabla k =  \tilde{\bX}^\ast \bS$ where $\bS$ is a diagonal $\tilde{m} \times \tilde{m}$ matrix. Indeed, a straightforward application of the chain rule shows that $\nabla k$ takes this form for any nonlinear combination of distance kernels and inner product kernels (defined in section \ref{Sec:designingKernels}).
Thus, for this extremely broad class of kernels, the linear operator may be expressed in the general form
\begin{align}
	\bL = \tilde{\bW} \tilde{\bX}^\ast \bS
	\label{Eq:linForm}
\end{align}
where $\bS$ is a diagonal matrix that depends only the choice of kernel and base state.

In the case $\tilde{m} \ll n$, the expression \eqref{Eq:linForm} is computationally attractive since $\bL$ need not be stored explicitly;
instead of storing a large $n\times n$ matrix it is sufficient to store two $n\times \tilde{m}$ matrices and a diagonal $\tilde{m} \times \tilde{m}$ matrix.
Additionally, the potentially expensive matrix multiplications involved in forming $\bL$ can be avoided in most practical scenarios.
For example, it is not necessary to form $\bL$ explicitly if all that is required is its eigendecomposition, as with DMD.

%This efficient eigendecomposition is explored below.

\subsection{Extracting structure from kernel machines: the dynamic mode decomposition} \label{Sec:DMD}
We can exploit the factorisation in \eqref{Eq:linForm} to perform a dynamic mode decomposition of the linear operator $\bL$.
This step can be computationally expensive as $\bL$ is an $n \times n$ matrix so the  eigendecomposition costs $\mathcal{O}(n^3)$ operations.
However, we can obtain the leading eigenvectors and eigenvalues by computing the eigendecomposition of a much smaller matrix that is (at most) $\tilde{m}\times \tilde{m}$.
This idea is formalised in the following lemma.
\begin{mylemma}[Dynamic mode decomposition of the linear operator]
	\label{lemmaModes}
Let $\bS\tilde{\bX} = \bUp \bSp \bVp^\ast$ be the (economy) singular value decomposition of the rescaled dictionary and ${\hat{\bL} = \bUp^\ast \bL \bUp}$ be the projection of $\bL$ from \eqref{Eq:linForm} onto the columns of $\bUp$.
	If $\hat{\bL} \hat{\bpsi} = \lambda \hat{\bpsi}$ with $\lambda \neq 0$ then
 \begin{align}
	 {\bpsi} = \frac{1}{\lambda}  
	 \tilde{\bW} \bVp
	 \bSp
	\hat{\bpsi}
	\label{Eq:eigen}
 \end{align}
is an eigenvector of $\bL$ with eigenvalue $\lambda$.
Additionally, all non-zero eigenvalues of $\bL$ are eigenvalues of $\hat{\bL}$.
\end{mylemma} %
The operator $\hat{\bL}$ represents the projection of the full linear operator $\bL$ onto the principal components 
(proper orthogonal decomposition modes) of the rescaled dictionary $\bS \tilde{\bX}$. 
These principal components can be computed via a batch SVD or computed online in parallel with the dictionary
and regression through a series of rank-one updates~\cite{Bunch1978,Brand2006a}.
%The lemma also holds if the SVD is replaced with a QR factorisation, but the SVD can be more easily
%truncated to obtain approximations to the modes.
This lemma is significant since it implies that every non-zero eigenvalue of $\bL$ can be obtained by computing the eigendecomposition of the smaller matrix $\hat{\bL}$.
Furthermore, the eigendecomposition produces an eigenvector of $\bL$ that corresponds to each eigenvalue.

We now prove the lemma using similar arguments to those used in theorem 1 of~\cite{Tu2014}.
\begin{proof}
	We first show that the pair $(\bpsi,\lambda)$ is indeed an eigenvector/eigenvalue pair.
	Assume that $\hat{\bL} \hat{\bpsi} = \lambda \hat{\bpsi}$ for $\lambda \neq 0$ and define
 \begin{align}
	 \bG = \tilde{\bW} \bVp\bSp
 \end{align}
 so that $\bpsi = \frac{1}{\lambda} \bG \hat{\bpsi}$.
 By \eqref{Eq:linForm} and the economy SVD of $\tilde{\bX}$,
 we may write $\bL$ as
\begin{align}
	\bL = \tilde{\bW} \left( \bUp \bSp \bVp^\ast \right)^\ast
	= \tilde{\bW} \bVp \bSp \bU^\ast
	= \bG \bUp^\ast.
\end{align}
Similarly,
\begin{align}
	\hat{\bL} = \bUp^\ast \left(\tilde{\bW}  \left( \bUp \bSp \bVp^\ast \right)^\ast \right) \bUp
	= \bUp^\ast \tilde{\bW} \bVp \bSigma = \bUp^\ast \bG.
\end{align}
Thus, 
\begin{align}
 	\bL \bpsi = \frac{1}{\lambda}
	\left(\bG \bUp^\ast\right) \left( \bG \hat{\bpsi} \right)
= \frac{1}{\lambda} \bG \hat{\bL}\hat{\bpsi} = \bG \hat{\bpsi} = \lambda \bpsi
 \end{align}
as required.

We will now prove that every non-zero eigenvalue of $\bL$ is also an eigenvalue of $\hat{\bL}$. 
Let $(\bpsi, \lambda)$ be an eigenvector/eigenvalue pair and define $\bu = \bUp^\ast \bpsi$.
Then
\begin{align}
	\hat{\bL} \bu = \bUp^\ast \bG \bU^\ast \bpsi = \bUp^\ast \bL \bpsi
	= \lambda \bUp^\ast \bpsi = \lambda \bu.
	\label{Eq:eig2}
\end{align}
Note also that $\bu$ is not the zero vector. If it were, then 
$\bL \bpsi = \bG \bUp^\ast \bpsi =\bG \bu = \mathbf{0}$ and therefore $\lambda = 0$ which
contradicts our assumption that $\lambda \neq 0$.
Combining this observation with \eqref{Eq:eig2} shows that $\lambda$ is also an
eigenvalue of $\hat{\bL}$.
\end{proof}

Pseudocode for extracting the linear operator and computing the dynamic mode decomposition is available in algorithm  \ref{Alg:batch}.
{In appendix \ref{Ap:DMDcomparison} we demonstrate that we recover the exact DMD formulation~\cite{Tu2014} in the special case of a linear kernel.}

\begin{algorithm}[t]
	\caption{Learning the model and analysing the linear component}
\label{Alg:batch}
\textbf{Inputs}:
	      data matrices $\bX$ and $\bY$,
	      kernel $k$,
	      dictionary tolerance $\nu$ \\
\textbf{Outputs}:
model $\bfun$, constant $\bc$, linear component $\bL$, nonlinear component $\bN$, eigenvectors $\bpsi$, and eigenvalues $\lambda$
  \begin{algorithmic}
\State Build the dictionary $\tilde{\bX}$ according to algorithm \ref{Alg:dictionary}
\State Solve $\argmin_{\tilde{W}} \| \bY - \tilde{\bW} \, k(\tilde{\bX},\bX)\|_F$ (for example, $\tilde{\bW} = \bY\, k(\tilde{\bX},\bX)^\dagger$)
 \State Define $\bS$ according to \eqref{Eq:linForm}
 \State Form the model as $\bfun(\bx) = \tilde{\bW}\, k(\tilde{\bX},\bx)$
  \State Form $\bc$, $\bL$, and $\bN$ according to section \ref{Sec:disambiguate} and a choice of base state
  \State Compute the eigendecomposition of $\hat{\bL}$ according to lemma \ref{lemmaModes}
  \State Form the eigenvectors $\bpsi$ and eigenvalues $\lambda$ according to \eqref{Eq:eigen}
 \end{algorithmic}
\end{algorithm}

\subsection{Extracting structure from kernel machines: querying nonlinear relationships}
\label{Sec:query}
The analysis of section \ref{Sec:disambiguate} showed 
that we can extract linear relationships from otherwise opaque kernel machines.
This section demonstrates that we can also extract specific nonlinear relationships between the input and output states.

Suppose that we know that the implicit feature space consists of a specific nonlinear scalar feature of interest labelled $\phi_j(\bx)$.
For example, we may be interested in the effect of quadratic interactions between two states: $\phi_j(\bx)=x_1 x_2$.
To `query' $\phi_j(\bx)$ is to determine the $n$-dimensional vector that represents the effect of the nonlinearity $\phi_j(\bx)$ on the elements of the output vector $\bfun(\bx)$. 
Without loss of generality, we can decompose the implicit feature vector $\bphi(\bx)$ into $\phi_j(\bx)$ and $\bphi^\prime(\bx)$, 
where $\bphi^\prime(\bx)$ is the original feature vector with the $\phi_j(\bx)$ element removed.
Applying \eqref{Eq:kernelExp} allows us to write
\begin{align*}
    \by = \bfun(\bx) = 
    \tilde{\bW}\left(
    (\phi_j({\tilde{\bX}}))^\ast \phi_j(\bx) + (\bphi^\prime(\tilde{\bX}))^\ast \bphi^\prime(\bx)
    \right).
\end{align*}
Thus, the effect of the features $\phi_j(\bx)$ on $\bfun(\bx)$ can be determined by simply reading off its coefficient as $\tilde{\bW} (\phi_j({\tilde{\bX}}))^\ast$. 
The result corresponds to the $j$-th column of the explicit $\bXi$ matrix of coefficients from \eqref{Eq:SimpleExpansion}. 

\subsection{Using partial knowledge of system physics to design kernels}
\label{Sec:designingKernels}
An informed choice of kernel is critical to the success of kernel machines.
Prior knowledge about the physical properties of a system can -- and should -- be considered when designing the kernel used for learning. 
This physical knowledge may include specific symmetries, invariants, and conservation laws that are known to exist in the system under consideration. 
Similarly, knowledge of the specific partial differential equation governing the dynamics, such as the Navier--Stokes equations for fluids or the nonlinear Schr\"{o}dinger equation for lasers, may inform candidate terms that should be captured by the kernel.  
The ability to enforce this type of partial knowledge of the physics is one of the key strengths of the SINDy regression~\cite{Loiseau2018}, resulting in more accurate and stable models with less training data.  
Moreover, in addition to enforcing known physics, it is possible to uncover these physical properties when they are unknown based on which kernel functions provide the best validated performance.  
For example, the success of symmetric kernels suggests a deeper underlying symmetry in the system that generated the training data.  
In the next section, we will also see that it is possible to test several kernels, and by choosing the kernel with the best validated performance gain insight into what terms might be present in the governing equations.

The choice of kernel has many different perspectives as outlined in chapter 13 of~\cite{Scholkopf2002};
the most useful perspective in this work is that the kernel defines the function space used by our model.
For example, the kernel chosen in section \ref{Sec:kernelExample} corresponded to the function space of quadratic monomials.
Thus, that kernel can be used to model systems that are dominated by quadratic interactions between the states.

There are several strategies that can be used to design suitable kernels for a given physical problem.
Useful references are chapter 13 of~\cite{Scholkopf2002} and chapter 3 of~\cite{Shawe-Taylor2004}.
Kernels can be combined to obtain new kernels, which
affords significant flexibility when constructing kernels for a given problem.
For example, the set of (Mercer) kernels forms a convex cone: for kernels $k_1$ and $k_2$, the conical combination
\begin{align}
	k(\bu,\bv) = \alpha_1^2 k_1(\bu,\bv) + \alpha_2^2 k_2(\bu,\bv)
	\label{Eq:cone}
\end{align}
%1
is also a kernel.
When two kernels are combined in this way, their feature space representations are scaled and stacked.
If the kernels $k_{1,2}$ induce feature spaces $\bphi_{1,2}$ then the feature space 
induced by $k$ in \eqref{Eq:cone} is $\begin{bsmallmatrix} {\alpha_1}\bphi_1 \\ {\alpha_2}\bphi_2 \end{bsmallmatrix}$.
This construction is useful when designing kernels for a given physical problem.
For example, we may know that a system is dominated by linear and cubic interactions between its states.
Thus, we may propose a kernel consisting of conical combinations of appropriate monomial kernels:
\begin{align}
	k(\bu, \bv) =  \alpha_1^2 \bu^T \bv + \alpha_3^2 (\bu^T \bv)^3.
\end{align}
This kernel induces a feature space consisting of purely linear and cubic terms:
\begin{align}
	\bphi(\bx) = \left[ {\alpha_1}x_1 \quad \cdots \quad  {\alpha_1} x_n \quad
	{\alpha_3} x_1^3 \quad
	\sqrt{3} \alpha_3 x_1^2 x_1
\quad \cdots \quad 
\sqrt{6} \alpha_3 x_{n-1} x_{n-2} x_{n-3}
\quad {\alpha_3} x_n^3 \right]^T.
\end{align}
The constants $\alpha_{1,2}$ represent the relative importance of the linear and cubic terms and can
be chosen through physical intuition or cross validation.

Another useful result is that kernels are closed under direct sums.
If $k_1:\mathcal{X}_1 \times \mathcal{X}_1 \rightarrow \mathbb{R}$ and 
$k_2:\mathcal{X}_2 \times \mathcal{X}_2 \rightarrow \mathbb{R}$ are kernels then
then their direct sum
\begin{align}
	\left(k_1 \oplus k_2 \right)(\bu,\bu^\prime,\bv,\bv^\prime) =
	k_1(\bu,\bv) + k_2(\bu^\prime,\bv^\prime)
	\label{Eq:directSum}
\end{align}
is a kernel on 
$(\mathcal{X}_1 \times \mathcal{X}_2 ) \times (\mathcal{X}_1 \times \mathcal{X}_2)$.
This fact can be exploited to design kernels where the inputs have different
meanings or known physics implies different governing laws for the different states.
For example, we could have a state space consisting of two types of measurements so
$\bx= \begin{bsmallmatrix} \bx^{(1)} \\ \bx^{(2)} \end{bsmallmatrix}$
where $\bx^{(j)}$ are ${n^{\tiny(j)}}$-dimensional vectors.
Suppose also that it is known that the system is governed by a linear response to $\bx^{(1)}$ and quadratic interactions
of $\bx^{(2)}$.
An appropriate kernel for our model would then be
\begin{align}
k(\bu,\bv) = \alpha_1^2 \left(\bu^{(1)T} \bv^{(1)}\right) + \alpha_2^2 \left(\bu^{(2)T} \bv^{(2)}\right)^2
\end{align}
which induces the feature space
\begin{align}
	\bphi(\bx) = \left[ {\alpha_1}x_1^{(1)} \quad \cdots \quad  {\alpha_1} x_{n^{(1)}}^{(1)} \quad
		{\alpha_2} \left(x_1^{(2)}\right)^2 \quad
		\sqrt{2} \alpha_2 x_1^{(2)} x_2^{(2)} \quad \cdots
	\quad {\alpha_2} \left(x_{n^{(2)}}^{(2)}\right)^2 \right]^T.
\end{align}
Thus far we have only explained how to design kernels for purely polynomial models.
Non-polynomial terms play an important role in many nonlinear systems~\citep{Benner2020}, and these can easily be incorporated into kernel design.
For example, $\bl$ may represent a vector of  pointwise trigonometric functions that we wish to incorporate into our feature space.
The corresponding kernel is simply $k(\bu, \bv) = \bl(\bu)^T \bl(\bv)$,
which can be combined with any other kernel to supplement the feature space with the non-polynomial nonlinearities $\bl$.

Two classes of kernels have received significant attention in applications.
Inner product kernels take the form $k(\bu, \bv) = \kappa(\bu^T \bv)$ where $\kappa$ is a scalar function.
Inhomogeneous polynomial kernels are inner product kernels that take the form
\begin{align}
	k(\bu, \bv) = (c + \bu^T \bv)^d
	\label{Eq:polyKernel}
\end{align}
where $c$ is a constant and $d \in \mathbb{N}$ is the degree of polynomial. These inhomogenous polynomial kernels are linear combinations of the monomial kernels in \eqref{Eq:monKernel}.
The special case $c=0$ and $d=1$ corresponds to a linear feature space.

Distance kernels are another important class of kernels and take the form \mbox{$k(\bu,\bv) = \kappa (\| \bu - \bv \|_2)$}. 
A popular example that is used in several applications is the Gaussian kernel, which is expressed as
\begin{align}
	k(\bu,\bv) = \exp \left(-\|\bu - \bv\|_2^2/(2 \sigma^2) \right)
	\label{Eq:gaussKernel}
\end{align}
where $\sigma$ is a constant.
The Gaussian kernel is a similarity measure of the state space, and the induced feature space is infinite dimensional~\citep{Steinwart2006}.

Supplementing the feature space with a constant can be achieved by combining a kernel with a constant, such as $k = \alpha_0^2 + \alpha_1^2 k_1$,
so that the feature space becomes $\begin{bsmallmatrix} {\alpha_0}\\ {\alpha_1}\bphi_1 \end{bsmallmatrix}$.
Additionally, pointwise products of kernels ($k = k_1 k_2$) are also kernels and 
the corresponding features are the products of all pairs of features from the first and second feature space.

Kernels can also be designed to respect known physical invariances or symmetries~\cite{Haasdonk2007}.
For example, Klus \emph{et al.}~\cite{Klus2021} recently derived analogues of the Gaussian and polynomial kernels that respect the symmetries of quantum physics.
Models that respect such invariances and symmetries are highly desirable as they usually require less training data and are less prone to overfitting.
Techniques for incorporating invariances into kernel machines are available in  chapter 11 of~\cite{Scholkopf2002} and~\cite{Decoste2002,Haasdonk2007}.

To summarise this section, we have demonstrated that%
\begin{enumerate}
	\item kernels can efficiently compute dense polynomial interactions between states
		that would otherwise be combinatorially complex,
	\item kernels can be combined to generate a range of feature spaces,
	\item if the features have different physical meanings or governing laws then one can 
		construct separate models and combine the kernels using a direct sum,
	\item non-polynomial and constant terms can be incorporated into kernels, and,
	\item kernels can be designed to respect symmetries and invariances.
\end{enumerate}
These observations indicate that there is significant flexibility for incorporating known partially known physics into our models through a suitable choice of kernel. 
Similarly, the validated performance of a handful of candidate kernels may provide insight into underlying physics, such as symmetries and terms in the governing equations.  
\section{Results}
\label{Sec:Applications}

\begin{figure}[t]
	\centering
	\includegraphics[width=\linewidth]{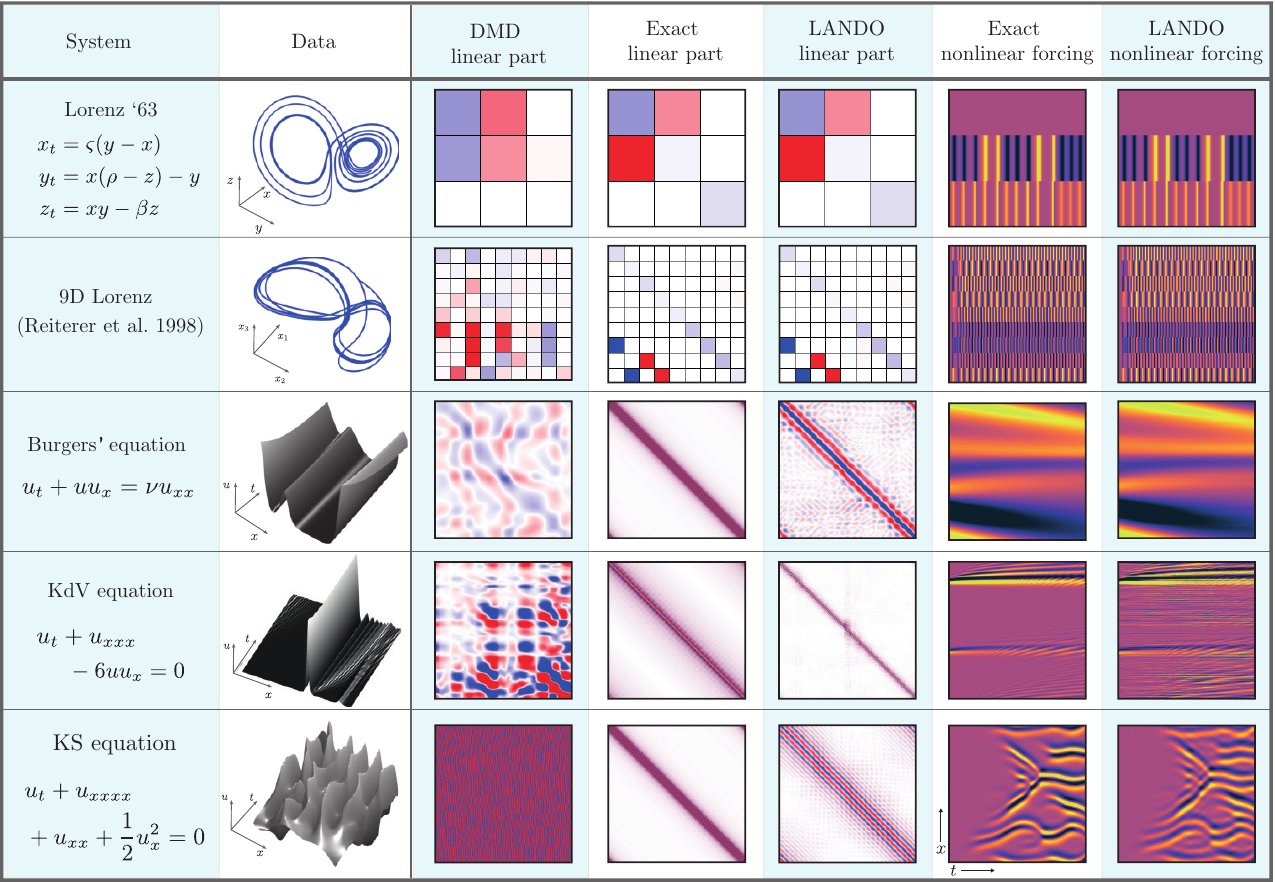}
	\caption{A comparison of learned linear operators for dynamical systems and PDEs.
	In the linear operators, red represents positive quantities whereas blue represents negative quantities.}
	\label{Fig:linearComparison}
\end{figure}

We now demonstrate our approach on a range of physically-relevant systems.
We will consider both dynamical systems and high-dimensional discretised PDEs.
The results of LANDO applied to these systems is summarized in Fig.~\ref{Fig:linearComparison}.
It can be seen that the true linear and nonlinear forcing components are accurately recovered by LANDO, while DMD fails to identify the correct linear model.  
For the dynamical systems considered, the linear operators are known exactly;
for the PDEs, we express the `true' linear operators as the appropriate spectral differentiation matrices.
In addition to the Lorenz system (\ref{Sec:Lorenz}), the viscous Burgers' equation (section \ref{Sec:Burgers})
and the KS equation (section \ref{Sec:KS}),
we also consider two other systems: the Korteweg–De Vries (KdV) equation for modeling waves on shallow water surfaces and a 9D analogue of the Lorenz system~\cite{Reiterer1998}. 
Reiterer \emph{et al.}  derived this analogue by modeling dissipative Rayleigh--Benard convection 
in a three-dimensional cell and applying a triple Fourier expansion to the associated Bousinnesq--Oberbeck equations. The resulting analogue exhibits similar asymptotic behaviour to the original Lorenz system, including a low-dimensional chaotic attractor and a period-doubling cascade.
We do not repeat the equations here for the sake of brevity, but they are analogous to the original Lorenz system and can be found in equation 18 of~\cite{Reiterer1998}.
In particular, the equations consist of a linear operator and quadratic interactions between the states.
We observe in figure \ref{Fig:linearComparison} that we recover the linear component of the operator to a high degree of accuracy.

\subsection{Implicit learning of the chaotic Lorenz system}
\label{Sec:Lorenz}
We first illustrate our approach on the Lorenz system~\citep{Lorenz1963}, which is a prototypical example of chaos and is often used in demonstrations of nonlinear system identification~\citep{Brunton2016}:
\begin{align}
\begin{split}
	\dot{x} &= \varsigma(y-x),\\
	\dot{y} &= x(\rho - z) - y, \\
	\dot{z} &= x y - \beta z,
	\end{split}
	\label{Eq:Lorenz}
\end{align}
where $\varsigma$, $\beta$, and $\rho$ are constants that parametrise the system.
Both the state dimension of the system ($n=3$) and the order of polynomial nonlinearity ($d=2$) are relatively small, so the benefits of kernel methods here are limited.
As such, the system is considered here only for demonstration.

We take the standard parameter values $\varsigma = 10$, $\rho = 28$, and $\beta = 8/3$ and initial condition $x = -8$, $y=8$, and $z=27$.
The system \eqref{Eq:Lorenz} is integrated from $t=0$ to $t=10$ and the solution is sampled at time intervals of $\Delta t = 10^{-3}$ resulting in 10,000 samples.
The data matrix $\bX$ comprises snapshots of the solution at each time step so that $\bx_j = [x(j \Delta t) \quad y(j\Delta t) \quad z(j\Delta t)]^T$,
and the columns of $\bY$ are the derivatives at each time:
$\by_j = \dot{\bx}_j$.
The order of the samples is randomly permuted so that the sparse dictionary is as rich as possible. 
The data use in this example are free of noise, and we demonstrate that the algorithm can be made robust to noise in appendix \ref{Sec:noise}.

\begin{figure}
    \begin{center}
    \includegraphics[width = \linewidth]{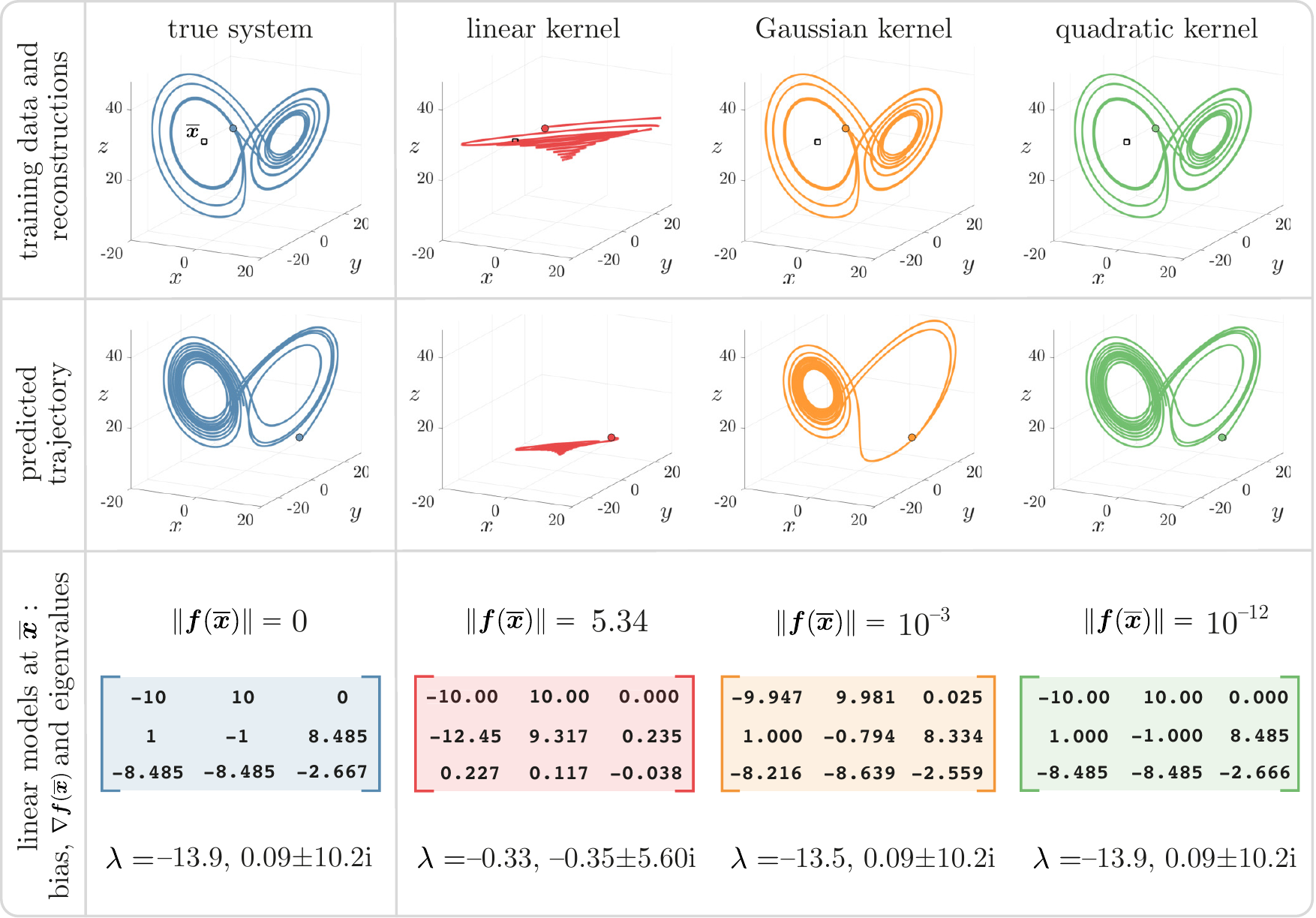}
    \end{center}
    \vspace{-.2in}
    \caption{Kernel learning of the Lorenz system.
    We compare the learned models and predicted trajectories for for linear, quadratic and Gaussian kernels.
    The training data are discrete time snapshots of the state~$[x\; y \; z]^T$ and the corresponding velocity measurements.
    The top row shows the models' reconstructions of the training data, the middle row shows the predicted trajectory from a different initial conditions, and
    the bottom row shows the learned linear model near the equilibrium point
   $\overline{\bx} = [ -\sqrt{\beta(\rho-1)} \quad - \sqrt{\beta(\rho-1)} \quad \rho - 1 ]^T$, which is indicated by $\square$.
    The parameter values are $\varsigma = 10$, $\rho = 28$ and $\beta = 8/3$ and the initial conditions are represented by $\circ$.
    }
    \label{Fig:Lorenz}
\end{figure}

The results of our kernel learning algorithm are illustrated in figure \ref{Fig:Lorenz}.
We consider three types of kernels: linear {($k(\bu,\bv) = \bu^T \bv$)}, quadratic 
{($k(\bu,\bv) = (1 + \bu^T \bv)^2$)},
and Gaussian  {($k(\bu,\bv) =\exp \left(-\|\bu - \bv\|_2^2/(2 \sigma^2) \right)$ with $\sigma = 1.1$)}.
These kernels are not optimised, and the best kernel parameters may be chosen through cross-validation.
The top row of figure \ref{Fig:Lorenz} illustrates the reconstructions achieved by each model on the same initial condition used for training.  
Each reconstruction is created by integrating the learned kernel model $\dot{\bx} = \bfun(\bx)$.
The linear model performs poorly and reconstructs a decaying spiral.
In contrast, the quadratic and Gaussian models accurately capture the behaviour of the underlying system.
The quadratic model has a training error of $\mathcal{O}(10^{-12})$.
Higher-order polynomial kernels produce similar training errors to the quadratic model.

This example also illustrates the value of a sparse dictionary: applying a standard kernel regression to this problem would require inverting a large $10,000\times 10,000$ matrix. Instead, the dictionary sizes are $\tilde{m} = 3,\, 7$, and $84$ {for the linear, quadratic, and Gaussian kernels}, respectively.

We also present the trajectories predicted by our models with the different initial condition of $x = 10$, $y=14$, and $z=10$.
The linear model prediction is poor and decays to the origin, while the Gaussian kernel model reproduces a trajectory that is rougly similar to the Lorenz system.
Finally, the quadratic kernel model generates an excellent prediction, which indicates the generalizability of the model to trajectories away from the initial data.
Given the form of the Lorenz system \eqref{Eq:Lorenz}, it is unsurprising that the quadratic kernel is the most accurate. However, the Gaussian kernel also performed well without any prior knowledge of the system form.
Thus, Gaussian kernels can be a good choice of model when partial knowledge of the underlying system is unavailable.

We now extract meaningful physical information from the kernel models.
Specifically, we extract linear models near the equilibrium
$\overline{\bx} = [ -\sqrt{\beta(\rho-1)} \quad - \sqrt{\beta(\rho-1)} \quad {\rho - 1} ]^T$, indicated by a square $\square$ in the top row of figure \ref{Fig:Lorenz}.
Note that $\overline{\bx}$ is not included in the training data.
Nevertheless, the quadratic and Gaussian models both identify $\overline{\bx}$ as an equilibrium since $\|\bfun(\overline{\bx}) \|$ is close to zero for both models.
Moreover, applying the results of section \ref{Sec:disambiguate} to each model generates local linear models for the behaviour near $\overline{\bx}$.
The true linearized model and the learned local linear models are reported in the third row of figure \ref{Fig:Lorenz}. 
All models capture the first row of the linearization, where the true system is also linear.  
However, the linear kernel model fails to estimate the rest of the linearization, while the quadratic and Gaussian kernel models provide excellent agreement; the local linear model learned by the quadratic kernel is correct to $\mathcal{O}(10^{-4})$.

\subsection{Extracting natural frequencies from densely coupled oscillators}
We now use our framework to study systems of coupled oscillators from a data-driven perspective.
The Kuramoto model is a prototypical model of coupling and synchronisation, and has been applied to biological, chemical, physical, and social systems~\cite{Acebron2005}.
We consider a forced Kuramoto model of $n$ coupled oscillators of the form
\begin{align}
   \dot{\vartheta} _{i} = \omega _{i}+\frac{1}{n} \sum _{j=1}^{n} a_{ij} \sin(\vartheta _{j}-\vartheta _{i}) + h \sin(\vartheta_j),\qquad i=1,\ldots, n,
   \label{Eq:sync}
\end{align}
where $\{\vartheta_i(t)\}$ are the phases, $\{ \omega_i\}$ are the natural frequencies, $h$ is a forcing constant, and $a_{i,j}$ are constants representing the nonlinear coupling between the $i$-th and $j$-th oscillators.
This example is inspired by similar recent studies of~\cite{Gelss2019} and~\cite{snyder2021data}, who sought
to learn predictive models for the Kuramoto system.
Instead, our aim here is to extract structural model information from the system.
In particular, we wish to learn the natural frequencies of each oscillator, $\omega_i$.

To train our model, we follow~\cite{Gelss2019} and use the kernel
\begin{align}
    k(\bu,\bv) = \left(c + \begin{bmatrix} \sin(\boldsymbol{\bu}) \\ \cos(\boldsymbol{\bu}) \end{bmatrix}^\ast \begin{bmatrix} \sin(\boldsymbol{\bv}) \\ \cos(\boldsymbol{\bv}) \end{bmatrix}\right)^2,
    \label{Eq:syncKernel}
\end{align}
producing a feature space consisting of constant, linear, and quadratic trigonometric terms.
This is an example of a kernel with non-polynomial terms, as was discussed in section \ref{Sec:designingKernels}.
We seek $\bfun$ that defines the dynamical system \ref{Eq:sync} such that $\dot{\bx} = \bfun(\bx)$.
The natural frequencies are the constant term in \eqref{Eq:sync} so, by section \ref{Sec:disambiguate}, the natural frequencies are approximated by $\bomega \approx \bfun(\boldsymbol{0})$.

\begin{figure}
    \begin{center}
    \includegraphics{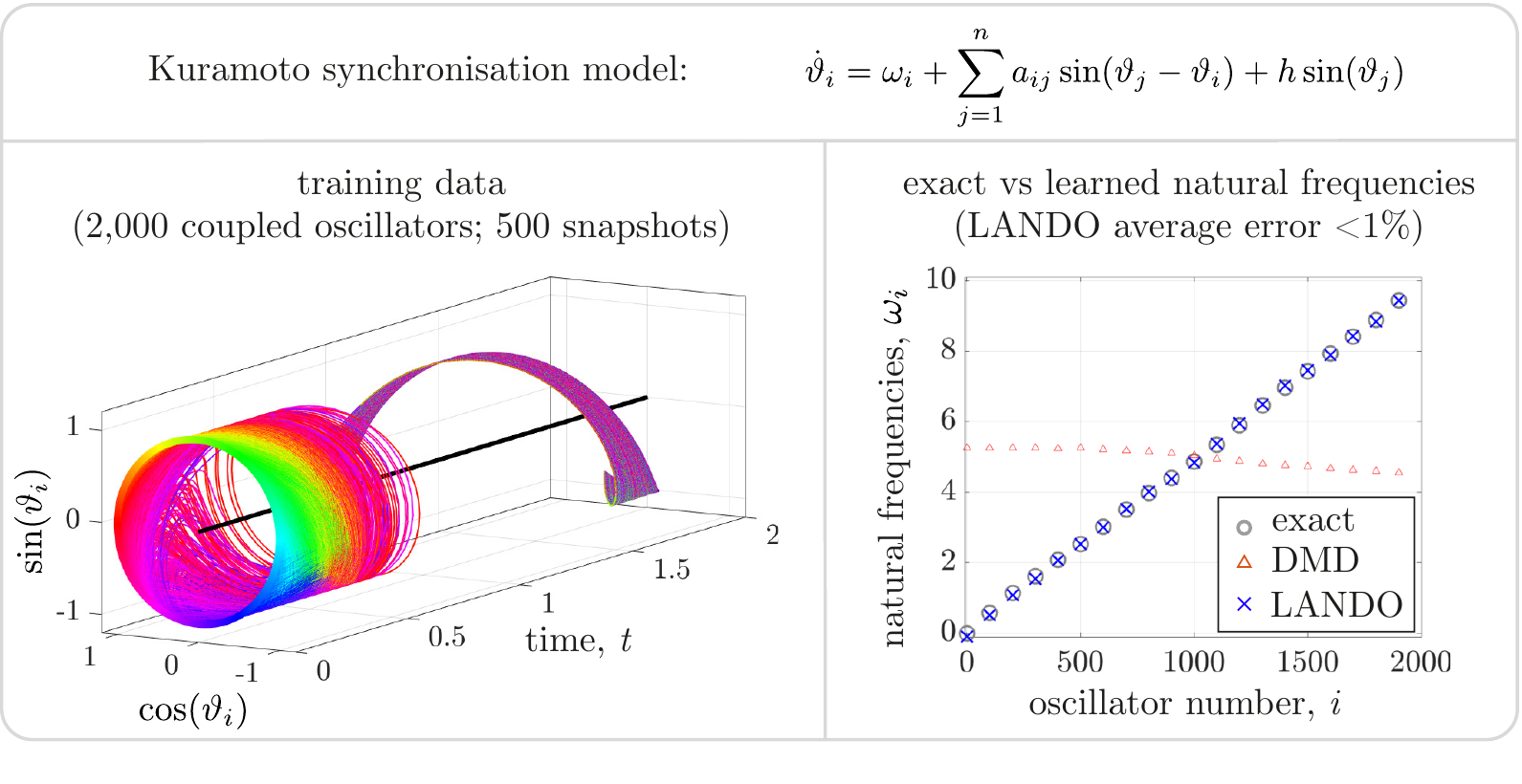}
    \vspace{-.2in}
    \caption{Learning the natural frequencies of coupled oscillators.  The training data is generated from a forced Kuramoto model and is illustrated on the left panel. The LANDO framework extracts the natural frequencies of the model. These learned natural frequencies are compared to the true natural frequencies on the right panel and the frequencies learned by a linear (DMD) model; because there are $2,000$ oscillators, only a handful of frequencies are plotted.}
        \label{Fig:sync}
    \end{center}
\end{figure}

We consider a system of 2,000 coupled oscillators with state vector $\bx =[\vartheta_1 , \, \cdots \, , \vartheta_{2,000} ]^T$.
The data is plotted in figure \ref{Fig:sync}. 
The feature space, which is implicitly defined by \eqref{Eq:syncKernel}, has over $2 \times 10^6$ elements, which is prohibitively expensive to work with explicitly.
We consider a strongly coupled system and randomly sample the coupling constants $a_{ij}$ from a normal distribution with mean 15 and variance 5. We take a forcing value of $h=2$, and randomly sample $\omega_i$ from a uniform distribution on the interval $[0,10]$.
The system is integrated to $t=2$, and we consider only a single simulation.

\subsection{Learning the spectrum of the viscous Burgers' equation}
\label{Sec:Burgers}
We now apply our learning framework to study a partial differential equation.
The Burgers' equation is a simplified version of the Navier--Stokes equations and is a prototypical nonlinear hyperbolic PDE.
The one-dimensional Burgers' equation takes the form
\begin{align}
	u_t
 =\nu u_{xx} - u u_x
 \label{Eq:Burgers}
 \end{align}
%  \begin{align}
%  x \in \left[-1, 1\right], \qquad \qquad u(-1) = u(1).
% \end{align}
%
where $u(x,t)$ is the velocity at position $x\in [-1,1]$ and time $t\geq 0$, and $\nu$ is the kinematic viscosity.

\begin{figure*}[t]
\vspace{-.45in}
	\centering
	\includegraphics[width = \linewidth]{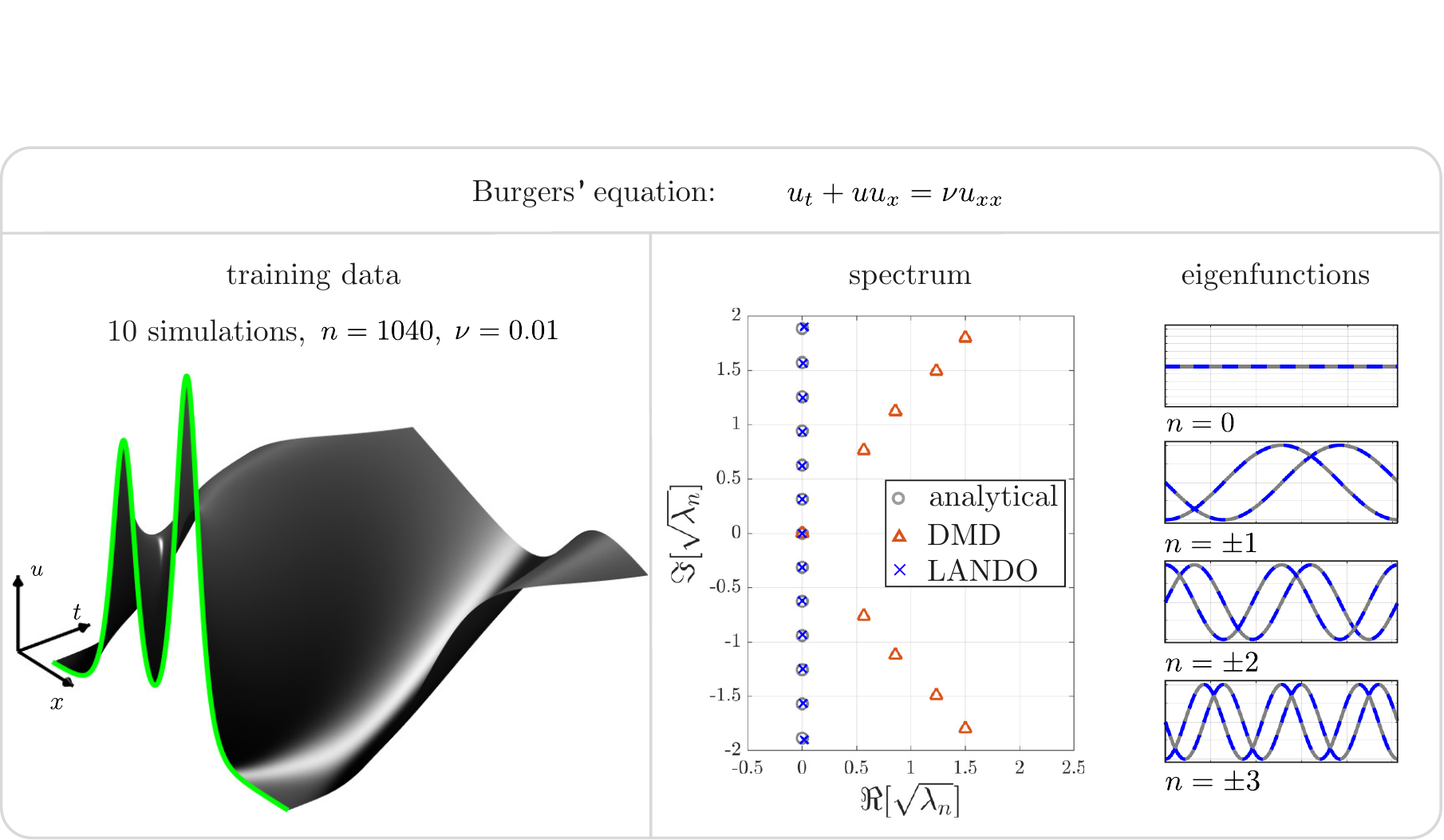}
	\vspace{-.05in}
	\caption{Learning the spectrum of the viscous Burgers' equation.
	A typical simulation is illustrated on the left with the initial condition highlighted in green. 
	The algorithm is trained on discrete time snapshots $\by_j = \bx_{j+1}$; velocity measurements $\dot{\bx}$ are not used in the training set.
	The figures on the right indicate that the algorithm accurately learns the eigenvalues and eigenfunctions of
	the linearised operator at the state $u\equiv0$.
	}%
	\label{Fig:burgers}
\end{figure*}

We simulate \eqref{Eq:Burgers} with periodic boundary conditions using the \texttt{spin} operator in the Chebfun package (\url{www.chebfun.org},~\cite{Chebfun}).
The solver uses exponential time differencing with fourth-order stiff time-stepping
\mbox{(ETDRK4,~\cite{Cox2002})};
a survey and comparison of these methods is available from~\cite{Montanelli2020}.
The same method is used to solve the other PDEs in this paper.
%Periodic boundary conditions are assumed and the initial conditions are specified for each problem.
The kinematic viscosity is $\nu = 0.01$ and we use initial conditions
\begin{align*}
 u(x,0) = 
3 A_{1} \textrm{sech}^2(3 \sin(\pi (x - 2 s_1)))
 +
5 A_{2} \textrm{sech}^2(3 \sin(\pi (x - 2 s_2)))
 %= (3*rand*sech(3*sin(pi*(x-2*rand)/2)).^2 + 5*rand*sech(3*sin(pi*(x-2*rand)/2)).^2);
\end{align*}
where $A_{1,2}$ and $s_{1,2}$ are constants randomly distributed in the interval $[0,1]$.
We perform 20 simulations and integrate to $t=1$;
a typical simulation is shown in the left panel of figure \ref{Fig:burgers}.

We train our model on the state vector defined by the solution $u$ sampled at spatial grid points separated by $\Delta x$
at time intervals of $\Delta t$ so that
\begin{align*}
 \bx_j = 
 \begin{bmatrix}
 u(-1,j \Delta t)
 &
 u(-1 + \Delta x,j \Delta t)
 & \dots
 %&u(1- \Delta x,j \Delta t)
 &u(1 - \Delta x,j \Delta t)
 \end{bmatrix}^T.
\end{align*}
We use 1024 spatial grid points and take $\Delta t = 10^{-3}$. 
We learn a discrete-time flow map that advances the state vector forward in time by $\Delta t$, so $\by_{j} = \bx_{j+1}$.
% Note that we do not use measurements of $\dot{\bx}_j$ in this example as these are typically unavailable in high-dimensional systems.
The data are uncorrupted by noise; learning the Burgers' equation in the presence of noise is explored in section~\ref{Sec:noise}.

We use a quadratic kernel to learn a model of this system, and from this model extract a local linearization relative to the equilibrium base state $\overline{\bx} = \boldsymbol{0}$.
The analytical linear operator is simply the Laplacian operator $\mathcal{A}u = \nu u_{xx}$. Since the boundary conditions are periodic, the eigenvalues are $\lambda_n = -\nu n^2 \pi^2$ for $n=0,1,2,\dots$.
All non-zero eigenvalues have multiplicity two and
the eigenfunctions are simply sines and cosines: {$\phi_n(x) = \sin(\lambda_n x),\; \cos(\lambda_n x)$}.

The analytical spectrum is compared to that learned by the kernel method in the right panels of figure \ref{Fig:burgers}.
The eigenvalues are plotted on a square-root scale so that their spacing is uniform, and the results are compared to the spectrum learned by exact DMD~\cite{Tu2014}.
The present algorithm accurately learns the true spectrum of the underlying linear operator whereas a na\"ive DMD implementation results in substantial errors.
The accuracy is best for eigenvalues with larger real part that are associated with slower dynamics, and deteriorates for the eigenvalues associated to quickly dampened modes.
Similarly, the kernel method accurately recovers the linear eigenfunctions.
The DMD eigenfunctions are very inaccurate and are therefore omitted from the figure. 

This example is particularly challenging, as indicated by the poor performance of DMD.
The choice of $\nu=0.01$ makes the effect of the linear operator $\nu \partial u_{xx}$ small compared to the nonlinear component $-u_x u$.
As such, it is particularly difficult for the algorithm to extract the underlying linear operator that is buried beneath nonlinear mechanisms.
Additionally, the choice of initial conditions did not provide a particularly rich set of data for the algorithm to work with.

%This example also illustrates the necessity of employing the dimensionality reduction techniques used in this paper.
%The data set was relatively large: if fully stored, the data matrices would be of the order of $10^3 \times 10^4$.

The relatively large size of the state space (${\sim}10^3$) and the high number of samples (${\sim}10^4$) emphasise the necessity of the dimensionality reduction techniques employed in this paper.
The kernel trick means that the quadratic feature space need not be constructed explicitly.
To do so would require $O(10^6)$ operations which would result in 
significant training costs and severely limit the speed of predictions.
Additionally, if we did not use a sparse dictionary then the kernel matrix would  have $\mathcal{O}(10^4)$ rows and columns, which becomes costly to invert.
The dictionary size of this system is approximately 100, which indicates that there are around 100 states that significantly contribute to the underlying dynamics in the high-dimensional feature space.
These states are then selected to form the basis of the dynamical model. 

In summary, we have applied our learning framework to uncover the linear operator of the nonlinear Burgers' equation purely from data measurements. 
This example demonstrates that the algorithm can be used to uncover linear structure highly nonlinear PDEs.
We now progress to a more challenging example.

\subsection{Learning the spectrum of the Kuramoto--Sivashinsky equation}
\label{Sec:KS}
The Kuramoto--Sivashinsky (KS) equation is a PDE that is used to model a range of physical phenomena including turbulence, chemical reactions and flame fronts.
The PDE is defined by
\begin{align*}
u_{t} =- u_{xx} - u_{xxxx} -\frac{1}{2} uu_{x}
\end{align*}
for $x \in [-L/2,L/2]$, periodic boundary conditions, and some given initial condition.
Similar to Burgers equation, the nonlinear term ($-u u_x/2$) transfers energy from low to high wavenumbers.
However, in the KS equation, the sign of the second derivative is reversed and is thus energy producing.
Combining these features with the stabilising influence of the fourth derivative generates highly complex, sometimes chaotic behaviour.
It has been described as the ``simplest chaotic PDE''~\cite{Brummitt2009} and therefore represents a useful test case for our algorithm.

\begin{figure*}[t]
	\centering
	\includegraphics[width = \linewidth]{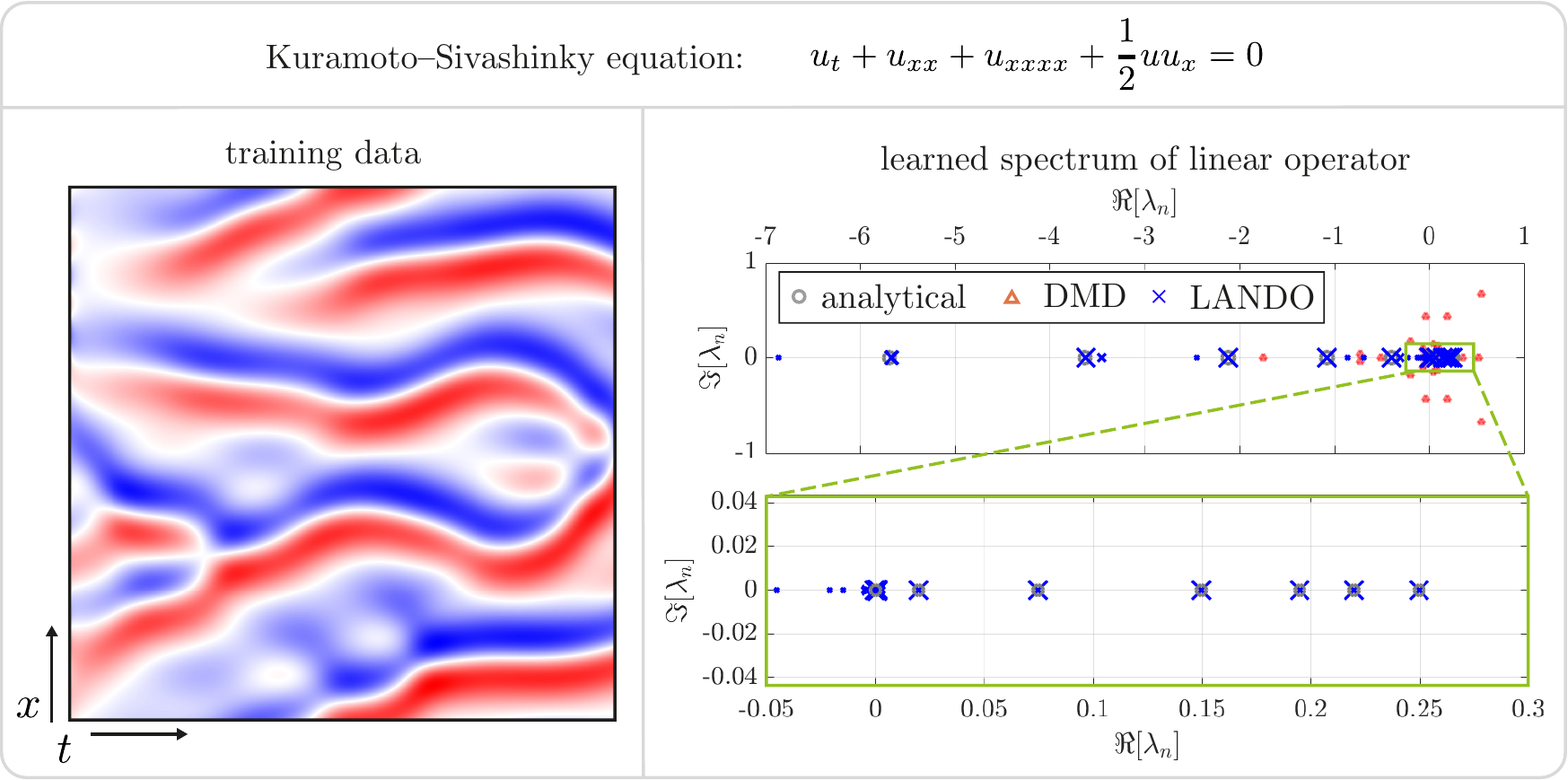}
	\vspace{-.275in}
	\caption{Learning the spectrum of the Kuramoto--Sivashinsky  equation.
	A typical simulation is illustrated on the left. 
	The algorithm is trained on discrete time snapshots $\by_j = \dot{\bx}_{j}$.
	The figures on the right indicate that the algorithm accurately learns the eigenvalues of
	the linearized operator at the state $u=0$.
The size of the markers of the LANDO eigenvalues correspond to the average projection of the training data onto the associated eigenvectors.
	}%
	\label{Fig:KS}
\end{figure*}

We use our kernel learning algorithm to recover the spectrum of the underlying linear operator relative to the equilibrium state $u=0$.
The linearized operator is 
$\mathcal{A}f = -f_{xx} - f_{xxxx}$.
Again, we consider periodic boundary conditions, so 
the eigenvalues are $\lambda_n = (n \pi)^2 - (n \pi)^4$ with
eigenfunctions $\phi_n= \sin(\lambda_n x)$, $\cos(\lambda_n x)$.
We simulate the KS equation using the ETDRK4 method described previously.
The initial conditions are now taken to be random periodic functions:
in particular, the initial conditions are finite Fourier series with distributed coefficients of equal variance~\cite{Filip2019}.
Our numerical experiments indicate that this rich set of initial conditions is necessary to ensure the algorithm's accuracy.
We define the box length as $L = 14\pi$, which is sufficiently large to generate chaotic behaviour.
A typical simulation is illustrated on the left panel of figure \ref{Fig:KS}.
The data matrices are constructed in a similar way to that of the Burgers' equation except we now use velocity measurements in the training data so $\by_j = \dot{\bx}_j$.
Again, we use 1024 spatial grid points and the samples are separated in time by $\Delta t = 0.05$.
We integrate the PDE to $t = 60$ and use 25 different simulations in the training data set.

The results of the learned spectrum are illustrated on the right panel of figure \ref{Fig:KS}.
The algorithm accurately learns the eigenvalues with the correct multiplicity.
Similarly to the Burgers' equation, the recovery of the smallest eigenvalues is most accurate, but the accuracy decreases for eigenvalues with smaller (more negative) real part.
The close-up figure also indicates that the  algorithm recovers the intricate behaviour of the spectrum for small eigenvalues.
Although they are not plotted, the eigenfunctions are also recovered to a high degree of accuracy.

This example demonstrates that our implicit learning method can extract accurate information about the underlying linear operator of a chaotic PDE.
This is performance is encouraging given our eventual goal of studying chaotic, turbulent fluid flows.
 
The right panel of figure \ref{Fig:sync} reports the results of the learned natural frequencies.
The average error of the estimates made by LANDO is less than 1\%. 
The deviations of the predictions are slightly worse at the upper and lower ends of the spectrum; the cause of this may be that the oscillators synchronise on the average natural frequency, which is $5$ here, and the model is more accurate for frequencies close to the average.
We also compare the results on a DMD model trained on linear combinations of $\sin(\boldsymbol{\vartheta})$, $\cos(\boldsymbol{\vartheta})$ and a constant vector. The learned natural frequencies of the linear model are very inaccurate, which illustrates the need of incorporating nonlinearities when attempting to learn the underlying natural frequencies.
\section{Discussion}%
\label{Sec:Discussion}
We have presented a data-driven kernel method that robustly extracts dynamic modes from high-dimensional, nonlinear data.
The method may be viewed as a confluence of the dynamic mode decomposition, the sparse identification of nonlinear dynamics, and kernel methods. 
Specifically, we use a kernelized identification of nonlinear dynamics (INDy, i.e., SINDy without the sparsity promoting regularizer) to robustly disambiguate linear and nonlinear dynamics, enabling the extraction of an explicit linear DMD model and forcing snapshot matrix. 
Access to the disambiguated DMD model and forcing snapshot matrix opens up the possibility of performing data-driven resolvent analysis of strongly nonlinear flows~\cite{herrmann2021jfm}. 
Our approach is based on the kernel recursive least squares algorithm~\cite{Engel2004} and kernel dynamic mode decomposition~\cite{Williams2015b} but introduces several innovations, including stabilised dictionary learning, improved interpretability, and extraction of locally-linear models and forcing.
We have demonstrated our approach on a range of nonlinear dynamical systems and PDEs, and shown in each case that we can effectively disambiguate the roles of linearity and nonlinearity.
The nature of kernel methods, along with the online learning variant, render our approach suitable for data that is high-dimensional in both space and time.

\subsection{Limitations of the method}
There is significant scope for modifications, improvements, and generalisations of our framework.
In this section we outline a few key issues.
The effects of noise are discussed in appendix \ref{Sec:noise}.

In section \ref{Sec:designingKernels} we provided principles for designing kernels that incorporate partially known physics.
However, kernels invariably include some parameters that must be tuned.
For example, the constant $c$ in the polynomial kernel \eqref{Eq:polyKernel} and the bandwidth $\sigma$ in
the Gaussian kernel \eqref{Eq:gaussKernel}.
These parameters can be chosen via cross validation, an optimisation routine, a hierarchical Bayesian framework~\cite{Schwaighofer2005}, or
the recently proposed kernel flows~\cite{Owhadi2019}.

Nonlinear system identification is typically data intensive, and our algorithm is no exception.
Experience indicates that learning an adequate approximation of the linear spectrum of a PDE usually requires a relatively large number of snapshots.
For example, when learning the viscous Burgers' equation we used a space-discretized grid of ${\sim}10^3$ points and $10^4$ snapshots.
%This is not an issue from a computational perspective as we have developed streaming variants of our algorithm.
The relatively large data requirements could be ameliorated by enforcing known physics
such as symmetries and invariances, as discussed in section \ref{Sec:designingKernels}.
Alternatively, compressed sensing and random sampling techniques could be employed~\cite{Rudy2017}.
Note that these issues do not relate to computational complexity or memory -- the sparse 
dictionary learning and online variant already address these issues~\cite{Engel2004} -- but instead concern the data required to accurately learn the underlying model.

% It is worth mentioning that our algorithm is not a unit consistent method:
% if the algorithm is run on data matrices $(\bX, \bY)$ and
% then on data matrices $(\bD \bX, \bY)$ where $\bD$ is an arbitrary diagonal matrix
% then the results may not be consistent same.
% This is true for two reasons: the first is because nonlinear similarity in feature space will be warped
% by rescaling of the data so elements included in the dictionary may be different.
% The second reason is that the Moore--Penrose pseudoinverse is not unit consistent unless
% the matrix is of full row rank, although a unit-consistent matrix inverse was recently proposed by~\cite{Uhlmann2018}.

The right choice of regulariser is essential to the success of any machine learning algorithm.
In this work we used sparse dictionary selection as a regulariser to address the challenges described in section~\ref{Sec:method}.
However, there are many opportunities to include additional or alternative regularisers within our framework.
For example, regularisation can be incorporated into the minimisation problem \eqref{Eq:problemDictionary} in a number of ways.
The simplest approaches involve modifying the pseudoinverse $k(\tilde{\bX},\bX)^\dagger$ in the solution \eqref{Eq:genSol} to incorporate Tikhonov regularisation~\cite{Golub2013} or truncated-SVD regularisation~\cite{Hansen1987}.
It is also possible to incorporate an $l_1$ regulariser as a surrogate for sparsity promotion using, for example, the kernel Lasso algorithm~\cite{Roth2004,Wang2007}.
However, sparsity in $\tilde{\bW}$ does not necessarily produce sparsity in the nonlinear feature space, so the physical appeal of such an approach is limited.

\subsection{Extensions and applications of the method}
In addition to the extensions outlined in the previous section (improving sensitivity to noise, alternative regularisers, learning kernel parameters),
there are many other possible generalisations and applications of our method.

This study began with the ultimate aim of performing resolvent analysis~\cite{Mckeon2010,Mckeon2017,jovanovic2005jfm} of turbulent flows from a purely data-driven perspective.  
Over the past decades, advances in numerical methods~\cite{aakervik2006pof,bagheri2009aiaa,monokrousos2010jfm,loiseau2019springer,ribeiro2020prf} and the growing availability of computational power have enabled analysis of the linearised Navier--Stokes equations for flows of increasing complexity~\cite{bagheri2009jfm,theofilis2011arfm,luchini2014ardm}. 
The authors recently proposed a `data-driven resolvent analysis'~\cite{herrmann2021jfm} based
on the dynamic mode decomposition, but this approach is currently only applicable for linear flows because strong nonlinearity corrupts the linear DMD model. 
By separating the roles of linearity and nonlinearity, the present work opens the door to data-driven resolvent analysis of nonlinear and actively controlled PDEs.
The ability to perform resolvent analysis in a completely equation-free and adjoint-free manner removes the need to have intrusive access to a numerical solver. 
We are currently pursuing this approach for low-dimensional PDEs, though we expect that significant modifications to our approach will be needed before we can consider fully turbulent flows.

Models that are constrained to respect known physics require fewer training samples and are more robust to noise.
Thus, an extension that incorporates physical laws into our framework -- such as symmetries, invariances, and bifurcations -- would be a significant improvement.
In section \ref{Sec:designingKernels} we explored incorporating partial physical knowledge to design the kernel $k$, but there is also scope to incorporate known physics into the weight matrix $\bW$.
For example, the solution may be restricted to a physically consistent structure via constrained least-squares~\cite{Loiseau2018}.
An alternative approach is to employ a physics-informed regulariser to penalise departures from known physics -- this idea has been used to great effect in physics-informed neural networks~\cite{Raissi2019,Raissi2020}.

The LANDO algorithm can be combined with the recently proposed lift and learn framework~\cite{Qian2020}, which uses prior knowledge of a system's governing equations to construct a coordinate mapping where the dynamics are quadratic.
These quadratic models can be efficiently represented with a quadratic kernel model, as demonstrated on the Burgers' and KS equations in sections \ref{Sec:Burgers} and \ref{Sec:KS}.
Since the kernel method scales well with high-dimensional nonlinearities, it is also possible to consider higher-order nonlinear functions of the mapped variables, which allows more complex models to be considered, and reduces the burden of choosing the correct coordinate mapping.
The stabilised dictionary learning step of this work could also reduce the computational cost of the original kernel DMD algorithm as well~\cite{Williams2015b}. It will be interesting to see whether the advances presented herein -- of feature-space dictionary learning and interpretability -- could be applied to improve computations of the spectrum of the Koopman operator.

The online updating procedure (see appendix \ref{Ap:online}, also~\cite{Engel2004}) could be modified for time-varying systems by including  weighting, windowing, and forgetting factors (chapter 4,~\cite{Liu2010}).
This approach has recently been developed for DMD~\citep{Zhang2019}.
One application of this approach is to resolvent-based flow control~\cite{yeh2019jfm}.
\section*{Acknowledgements} 
This work was supported by U.S. Army Research Office (ARO W911NF-17-1-0306 and ARO W911NF-19-1-0045), the U.S. Office of Naval Research (ONR N00014-17-1-3022) and by the PRIME programme of the German Academic Exchange Service (DAAD) with funds from the German Federal Ministry of Education and Research (BMBF). SLB would like to thank Bing Brunton, Nathan Kutz, Jean-Christophe Loiseau, and Isabel Scherl for valuable discussions. 
\appendix
\section{Glossary of terms}
Table~\ref{Tab:Nomenclature} provides the nomenclature used throughout the paper.  
We have attempted to maintain consistency with DMD~\cite{Tu2014,Kutz2016}, SINDy~\cite{Brunton2016}, and KRLS~\cite{Engel2004} where possible, although several changes were made to unify the notation.  
Importantly, the features are denoted by $\bphi$ in this work, whereas they are denoted by $\btheta$ in SINDy. 
Similarly, the kernel weights are denoted by $\bW$ in this work, whereas they are denoted by $\bTheta$ in KRLS.  
\begin{table}[h]
	\def\arraystretch{1.3}%  1 is the default, change whatever you need
	\centering
\caption{A summary of terms used in the paper.}\label{Tab:Nomenclature}
\rowcolors{2}{white}{gray!15}
\begin{tabular}{ccL}
	  Symbol & Type & Meaning\\
    \cmidrule[0.9pt](lr{0.125em}){1-1}%
    \cmidrule[0.9pt](lr{0.125em}){2-2}%
    \cmidrule[0.9pt](lr{0.125em}){3-3}%
	  $n$ & number & State dimension \\
	  $m$ & number & Total number of samples \\
	  $N$ & number & Dimension of nonlinear feature space\\
	  $\bx$ & $n$-vector & State vector\\
	  $\by$ & $n$-vector & Model output (e.g. $\by = \dot{\bx}$ or $\by_k=\bx_{k+1}$)\\
	  $\bX$ & $n \times m $ matrix & Right data matrix \\
	  $\bY$ & $n \times m $ matrix & Left data matrix \\
	  $\bF(\bx)$ & function & Underlying system \\
	  $\bfun(\bx)$ & function & Approximate learned model \\
	  $\bc$ & $n$-vector  & Constant shift in model\\
	 % $\hat{\bA}$ & $r\times r$ matrix & Projection of $\bA$ onto POD modes\\
	  $\bL$ & $n\times n$ matrix & Linear component of true operator\\
	  %$\hat{\bL}$ & $r \times r$ matrix & Projection of $\bL$ onto POD modes\\
	  $\bN$ & function & Purely nonlinear component of operator\\
	  $\bA$ & $n\times n$ matrix & DMD best-fit operator\\ 
	  $k(\cdot, \cdot)$ & function & Kernel function\\
	  $\bphi(\cdot)$ & function & Nonlinear (implicit) feature space\\
	  $t$ & continuous variable & time \\
	  	  $t$ & discrete index & Snapshot number\\
	  	  $\nu$ & number & Dictionary sparsification parameter\\
  	  $\tilde{}$ & tilde & Indicates quantity is connected to the dictionary \\
	  $ \tilde{m}_t$ & number & Number of samples in dictionary at time $t$\\
	  $\tilde{\bK}_t$ & $\tilde{m}_t \times \tilde{m}_t$ matrix & Kernel matrix of dictionary elements at time $t$ \\
	  $\tilde{\chol}_t$ & $\tilde{m}_t \times \tilde{m}_t$ matrix & Cholesky decomposition of $\tilde{\bK}_t$\\
	  $ \tilde{m}$ & number & Final number of samples in the dictionary\\
	  $ \bPi_t$ & $\tilde{m}_{t} \times t$ matrix& Matrix that approximately maps samples before time $t$ onto the dictionary \\
	  $\hat{}$ & hat & Indicates quantity is projected onto POD subspace \\
	    	  $\overline{\bx}$ & $n$-vector  & Base state (e.g. statistical mean or equilibrium solution)
\end{tabular}
\end{table}
\section{Online learning variant}
\label{Ap:online}
In this section we derive the rank-one update equations used in the online regression algorithm.
The derivation is equivalent to that presented in \cite{Engel2004} except we consider vector-valued outputs and do not apply the inverted kernel matrix explicitly.
A summary of the procedure may be found in the pseudocode in algorithm \ref{Alg:online}.

The minimisation problem, without regularisation, at time $t$ is
\begin{align}
	\argmin_{\tilde{\bW}_t} \left\| \bY_t - \tilde{\bW}_t \,\,k(\tilde{\bX}_t,\bX_t) \right\|^2_F
	=
	\argmin_{\tilde{\bW}_t} \left\| \bY_t - \tilde{\bW}_t \tilde{\bPhi}_t^\ast\bPhi_t \right\|^2_F.
\end{align}
The representation \eqref{Eq:dictApprox} allows us to approximate the above as
\begin{align}
	\argmin_{\tilde{\bW}_t} \left\| \bY_t -  
	\tilde{\bW}_t \tilde{\bPhi}^\ast_t\tilde{\bPhi}_t \bPi_t\right\|_F^2
	=\argmin_{\tilde{\bW}_t} \left\| \bY_t -  
	\tilde{\bW}_t \tilde{\bK}_t \bPi_t \right\|_F^2.
\end{align}
The minimiser of the above is 
\begin{align}
	\tilde{\bW}_t  = \bY_t \left( \tilde{\bK}_t \bPi_t \right)^\dagger
	= \bY_t \bPi_t^\dagger \tilde{\bK}_t^\dagger
	=\bY_t \bPi_t^\ast
	\left( \bPi_t \bPi_t^\ast\right)^{-1} \tilde{\bK}_t^{-1}.
	\label{Eq:alph}
\end{align}
In the above, we have used the fact that, by construction, $\tilde{\bK}_t$ has full column rank and $\bPi_t$ has full row rank.

When a new sample is considered, it falls into two cases as outlined in section \ref{Sec:dictionary}.
Either the sample is almost linearly dependent on the current dictionary elements, or it is not.
The updating equations are different in each case and we derive them below.
In what follows, it is convenient to define
\begin{align}
	\bP_t = \left( \bPi_t \bPi_t^\ast\right)^{-1}
\end{align}
and
\begin{align}
	\bh_t = \frac{\bpi_t^\ast \bP_{t-1}}{1 + \bpi_t^\ast \bP_{t-1} \bpi_t}.
	\label{Eq:qUpdateALD}
\end{align}
\subsection*{Case I: Almost linearly dependent}
If the new sample is almost linearly dependent on the dictionary
elements then the dictionary is not updated: $\mathcal{D}_t = \mathcal{D}_{t-1}$.
Since the dictionary doesn't change, neither does the kernel matrix so
$\tilde{\bK}_{t} = \tilde{\bK}_{t-1}$.
The update rule for $\bPi$ is simply
\begin{align}
	\bPi_t = \begin{bmatrix}
		\bPi_{t-1} &
		\bpi_t
		\end{bmatrix}.
		\label{Eq:piUpdateALD}
\end{align}
Thus, $\bPi_t \bPi_t^\ast = \bPi_{t-1} \bPi_{t-1}^\ast + \bpi_t \bpi^\ast_t$, which corresponds to a rank-1 update.
Accordingly, the matrix inversion lemma says that the update rule for $\bP_t$ is
\begin{align}
	\bP_t = \bP_{t-1} - \frac{\bP_{t-1} \bpi_t \bpi_t^\ast \bP_{t-1}}
	{1 + \bpi_t^\ast \bP_{t-1} \bpi_t}
	= \bP_{t-1} - \bP_{t-1} \bpi_t \bh_t.
	\label{Eq:pUpdateALD}
\end{align}
We may now define the update rule for $\tilde{\bW}_t$.
Since
\begin{align}
	\bY_t \bPi_t^\ast = \bY_{t-1} \bPi_{t-1}^\ast + \by_t \bpi_t^\ast,
\end{align}
applying \eqref{Eq:pUpdateALD} to \eqref{Eq:alph} produces
\begin{align}
	\tilde{\bW}_t= \bY_t \bPi_t^\ast \bP_t\tilde{\bK}_t^{-1}
	  = \left( \bY_{t-1} \bPi_{t-1}^\ast + \by_t \bpi_t^\ast \right)
	  \left( \bP_{t-1} -\bP_{t-1} \bpi_t\bh_t \right) \tilde{\bK}^{-1}_t.
\end{align}
Expanding the brackets yields
\begin{align}
	\tilde{\bW}_t &= 
	\left(\bY_{t-1} \bPi_{t-1}^\ast \bP_{t-1}
	-\bY_{t-1}\bPi_{t-1}^\ast \bP_{t-1}\bpi_t \bh_t 
+ \by_t \bpi_t^\ast \bP_t\right)
\tilde{\bK}^{-1}_t.
\label{Eq:kern3}
\end{align}
Since we are not adding an element to the dictionary, the kernel matrix and its inverse remain the same:
$\tilde{\bK}_t^{-1}=\tilde{\bK}_{t-1}^{-1}$.
Additionally, from the regression in the previous iteration we have $\tilde{\bW}_{t-1} 
= \bY_{t-1} \bPi_{t-1}^\ast \bP_{t-1} \tilde{\bK}_{t-1}^{-1}$.
Thus, \eqref{Eq:kern3} can be expressed as
\begin{align}
	\tilde{\bW}_t &= 
	\tilde{\bW}_{t-1}+
	\left(
	 \by_t \bpi_t^\ast \bP_t	-\tilde{\bW}_{t-1} \tilde{\bK}_{t-1} \bpi_t \bh_t \right)
\tilde{\bK}^{-1}_t.
\end{align}
Finally, on use of $\bh_t = \bpi_t^\ast \bP_t$ and \eqref{Eq:smallPiUpdate}, the update rule for $\tilde{\bW}_t$ is
\begin{align}
	\tilde{\bW}_t &= \tilde{\bW}_{t-1} +
	\left( \by_t - \tilde{\bW}_{t-1} \tilde{\bk}_{t-1} \right)
	\bh_t \tilde{\bK}_t^{-1}.
	\label{Eq:thetaUpdateALD}
\end{align}
As discussed in section \ref{Sec:dictionary}, the product $\bh_t \tilde{\bK}_t^{-1}$
should be computed with two backsubstitutions with the Cholesky factor $\chol_t$.
\subsection*{Case II: Not almost linearly dependent}
In this case, the new vector is not almost linearly dependent.
Accordingly, we must add $\bx_t$ to the dictionary:
$\mathcal{D}_t= \mathcal{D}_{t-1} \cup \{\bx_t \}$.

The update rules for $\bPi_t$ and $\bP_t$ are simply
\begin{align}
    	\bPi_t =
	\begin{bmatrix}
		\bPi_{t-1} & \boldsymbol{0}\\
		\boldsymbol{0} & 1
	\end{bmatrix} \qquad \text{and}\qquad	\bP_t =
	\begin{bmatrix}
		\bP_{t-1} & \boldsymbol{0}\\
		\boldsymbol{0} & 1
	\end{bmatrix}.
	\label{Eq:pUpdateNALD}
\end{align}
The update rule for $\tilde{\bW}_t$ is
\begin{align}
	\tilde{\bW}_t= \bY_t \bPi_t^\ast \bP_t\tilde{\bK}_t^{-1}
			 = 
			 \begin{bmatrix}
				 \bY_{t-1} \bPi_{t-1}^\ast \bP_{t-1} \,\,&\,\,
				 \by_t
			 \end{bmatrix}
			 \tilde{\bK}_t^{-1}.
\end{align}
Finally, \eqref{Eq:Kupd2} allows us to express the update rule as
\begin{align}
	\tilde{\bW}_t &= 
	\begin{bmatrix}
		\tilde{\bW}_{t-1} + 
		\left( \tilde{\bW}_{t-1} \tilde{\bk}_{t-1} - \by_t \right)
		\dfrac{\bpi_t^\ast}{\delta_t} 
		\,\,\,\, &\,\,\,\,
		\left(  \tilde{\bW}_{t-1} \tilde{\bk}_{t-1} - \by_t \right)
		\dfrac{1}{\delta_t}
	\end{bmatrix}.
	\label{Eq:thetaUpdateNALD}
\end{align}
This completes the derivation of the equations for the online regression algorithm.
\begin{algorithm}[t]
	\caption{Online learning algorithm\\
		The operation count for each step is included on the right
	}
\label{Alg:online}
\textbf{Inputs}:
	      data matrices $\bX$ and $\bY$,
	      kernel $k$,
	      dictionary tolerance $\nu$\\
\textbf{Outputs}:
model $\bfun$, constant $\bc$, linear component $\bL$, 
nonlinear component $\bN$, eigenvectors $\bpsi$, and eigenvalues $\lambda$
  \begin{algorithmic}
  \For{$t = 1 \to m$}
  \State Select new sample pair $(\bx_t, \by_t)$
  \State Compute $\delta_t$ according to algorithm \ref{Alg:dictionary}
  \State Compute $\bpi_t$ using \eqref{Eq:smallPiUpdate}
  \Comment{$\mathcal{O}(\textcolor{col1}{\tilde{m}_t^2})$}
  \If{$\delta_t \leq \nu$ (almost linearly dependent)}
  \State Maintain the dictionary: $\mathcal{D}_t = \mathcal{D}_{t-1}$
  \State Update $\tilde{\bW}_t$ using \eqref{Eq:thetaUpdateALD}
  \Comment{$\mathcal{O}(\textcolor{col2}{n \tilde{m}_t})$}
  \State Compute $\bh_t$ using \eqref{Eq:qUpdateALD} 
  \Comment{$\mathcal{O}(\textcolor{col1}{\tilde{m}_t^2})$}
  \State Update $\bP_t$ using \eqref{Eq:pUpdateALD}
  \Comment{$\mathcal{O}(\textcolor{col1}{\tilde{m}_t^2})$}
  \ElsIf{$\delta_t > \nu$ (not almost linearly dependent)}
  \State Update the dictionary: $\mathcal{D}_t = \mathcal{D}_{t-1} \cup \{\bx_t\}$
  \State Update the Cholesky factor $\chol_t$ using \eqref{Eq:cholUpdate}
  \Comment{$\mathcal{O}(\textcolor{col1}{\tilde{m}_t^2})$}
  \State Update $\bP_t$ using \eqref{Eq:pUpdateNALD}
  \Comment{$\mathcal{O}(\textcolor{col0}{\tilde{m}_t})$}
  \State Update $\tilde{\bW}_t$ using \eqref{Eq:thetaUpdateNALD}
  \Comment{$\mathcal{O}(\textcolor{col2}{n \tilde{m}_t})$}
  \State Form the model $\bfun$ according to \eqref{Eq:fApprox}
\EndIf
  \EndFor
  \State Define $\bS$ according to \eqref{Eq:linForm}
    \State Form $\bL$, $\bc$ and $\bN$ according to section \ref{Sec:disambiguate}
  \State Compute the eigendecomposition of $\hat{\bL}$ according to lemma \ref{lemmaModes}
  \State Form the eigenvectors $\bpsi$ and eigenvalues $\lambda$ according to \eqref{Eq:eigen}
  \end{algorithmic}
\end{algorithm}

\section{Learning control laws}
\label{Ap:control}
Our algorithm may also be used to simultaneously learn control laws and governing equations from data.
An active control variable can significantly alter the behaviour of a dynamical system and thus further disguise the underlying dynamics~\cite{Proctor2016,kaiser2017arxivb}.
In many practical scenarios -- such as epidemiological modeling of disease spread where the control variables could be the distribution of vaccinations -- it is infeasible to gather data on the unforced system so the effects of control must be disambiguated from the data~\cite{proctor2015ih}.
This strategy can be used to uncover the dynamics of the unforced system,  which can then inform design of effective control strategies.
In such systems the underlying dynamics take the form
\begin{align}
	\by = \bF(\bx,\bu)
	\label{Eq:control1}
\end{align}
where $\bu$ is the control variable.
Similarly to $\bx$ and $\by$, the values of $\bu$ are known at each snapshot time.
As in dynamic mode decomposition with control (DMDc, \cite{Proctor2016}), we may
write the supplemented state vector as $\bomega = \begin{bsmallmatrix} \bx \\ \bu\end{bsmallmatrix}$
so that \eqref{Eq:control1} may be expressed as
\begin{align}
	\by = \bG(\bomega).
\end{align}
Our task is now to learn a model $\bg$ that approximates the underlying system defined by $\bG$.

% Designing an appropriate kernel for this model is key.
We can use ideas explained in section \ref{Sec:designingKernels}
to design a suitable kernel.
In particular, we can exploit the fact that kernels are closed under direct sums \eqref{Eq:directSum}.
Unless we believe that there are nonlinear pairings between the control variable and the state space, we can assume a kernel of the form
\begin{align}
	k(\bomega, \bomega^\prime) = k_x(\bx,\bx^\prime) + k_u(\bu,\bu^\prime).
	\label{Eq:controlKernel}
\end{align}
For example, we may have reason to believe that $\by$ is generated by 
quadratic interactions between the states $\bx$ whereas the control variable has only a linear effect.
Then, the kernel
\begin{align}
	k(\bomega, \bomega^\prime) = (\bx^\ast \bx^\prime)^2 + (\bu^\ast \bu^\prime)
\end{align}
induces the appropriate feature space.
The algorithm of section \ref{Sec:method} can then be applied with $\bX$ replaced by the augmented data matrix ${\bOmega} = \begin{bsmallmatrix} {\bX} \\ {\bU} \end{bsmallmatrix}$ to learn a model of the form
\begin{align}
	\bg(\bomega) = \tilde{\bW} \,\,k(\tilde{\bOmega},\bomega)
\end{align}
where $\tilde{\bOmega}$ is the augmented dictionary matrix.
If the kernel takes the form \eqref{Eq:controlKernel} then the reconstruction/prediction model is
\begin{align}
	\bg(\bomega) =  \tilde{\bW}_x k_x(\tilde{\bX},\bx)
	+\tilde{\bW}_u k_u(\tilde{\bU},\bu)
	\label{Eq:controlModel}
\end{align}
where $\tilde{\bOmega} = \begin{bsmallmatrix} \tilde{\bX} \\ \tilde{\bU} \end{bsmallmatrix}$ and $\tilde{\bW} = \begin{bsmallmatrix} \tilde{\bW}_x \,\, \tilde{\bW}_u \end{bsmallmatrix}$.
The unforced system can then be modeled by setting ${\tilde{\bW}_u = \boldsymbol{0}}$.
Furthermore, we can also compute local linear models (i.e., DMD models) of the unforced system using the ideas of section \ref{Sec:disambiguate}.
The analysis follows exactly except $\tilde{\bW} \, k(\tilde{\bX},\bx)$ in section \ref{Sec:disambiguate} is replaced with
$\tilde{\bW}_x \, k_x(\tilde{\bX},\bx)$.
Note that if the kernel is taken to be
\begin{align}
	k(\bomega, \bomega^\prime) =
	\bomega^\ast \bomega^\prime 
	= (\bx^\ast \bx^\prime) + (\bu^\ast \bu^\prime)
\end{align}
then we recover the original DMDc formulation \cite{Proctor2016}.

These ideas are also valid for kernels that don't take the form \eqref{Eq:controlKernel}
but the algebra is slightly more involved and we therefore do not report the results here.

\section{Connection to exact dynamic mode decomposition}
\label{Ap:DMDcomparison}
We now elucidate the connection between the present work
and exact dynamic mode decomposition of \cite{Tu2014}
which was introduced in section \ref{Sec:exactDMD}.
In particular, we demonstrate that exact DMD can be viewed as a special case of the present work when there
is no sparsity promotion and the kernel is linear.
The linear kernel is \mbox{$k(\bu,\bv) = \bu^\ast \bv$}, 
so the implicit feature space is simply $\bphi(\bx) = \bx$.
As such, the full model is the linear map $\bfun(\bx) = \bL \bx$ where $\bL$ is \eqref{Eq:linForm}
\begin{align}
	\bL = \tilde{\bW} \tilde{\bX}^\ast.
\end{align}
In exact DMD there is no sparsity promotion so the dictionary used in our algorithm is full:
$\bPi = \bI$ and $\tilde{\bX} = \bX$.
Moreover, $\tilde{\bW}$ is given by \eqref{Eq:fullSol} so
\begin{align}
	\bL = \bY \,\, k(\bX, \bX)^\dagger \,\, \bX^\ast.
\end{align}
Expanding the kernel yields
\begin{align}
	\bL = \bY \left(\bX^\ast \bX \right)^\dagger \bX^\ast
	= \bY \bX^\dagger (\bX^\ast)^\dagger \bX^\ast
	= \bY \bX^\dagger
	\label{Eq:finComp}
\end{align}
which is identical to the linear operator $\bA$ from \eqref{Eq:exactDMD} defined by exact DMD.
In \eqref{Eq:finComp} we used the identities for the Moore--Penrose pseudoinverse
$(\bM^\ast \bM)^\dagger = \bM^\dagger (\bM^\ast)^\dagger$ and 
$\bM^\dagger (\bM^\ast)^\dagger \bM^\ast = \bM^\dagger$ for any matrix $\bM$.

Similarly, the eigenmodes computed by exact DMD are equivalent to those defined in lemma \ref{lemmaModes} in the special case
of a linear kernel without sparsity promotion.
%The eigenmodes in exact DMD are \cite{Tu2014}
%%
%\begin{align}
%	\bw_{\rm exact} = \frac{1}{\lambda_{\rm exact}} \bY \bV \bSigma^{-1} \hat{\bw}_{\rm exact}
%\end{align}
%%
%where $\bX = \bU \bSigma \bV^\ast$ is the singular value decomposition of $\bX$
%and $(\hat{\bw}_{\rm exact}, \lambda)$ is an eigen-pair of the linear operator projected onto the POD modes:
%%
%\begin{align}
%	\hat{\bA}_{\rm exact} \hat{\bw}_{\rm exact} = \lambda_{\rm exact} \hat{\bw}_{\rm exact}.
%\end{align}
%

\section{Sensitivity to noise}
\label{Sec:noise}

\begin{figure*}[t]
	\centering
	\includegraphics{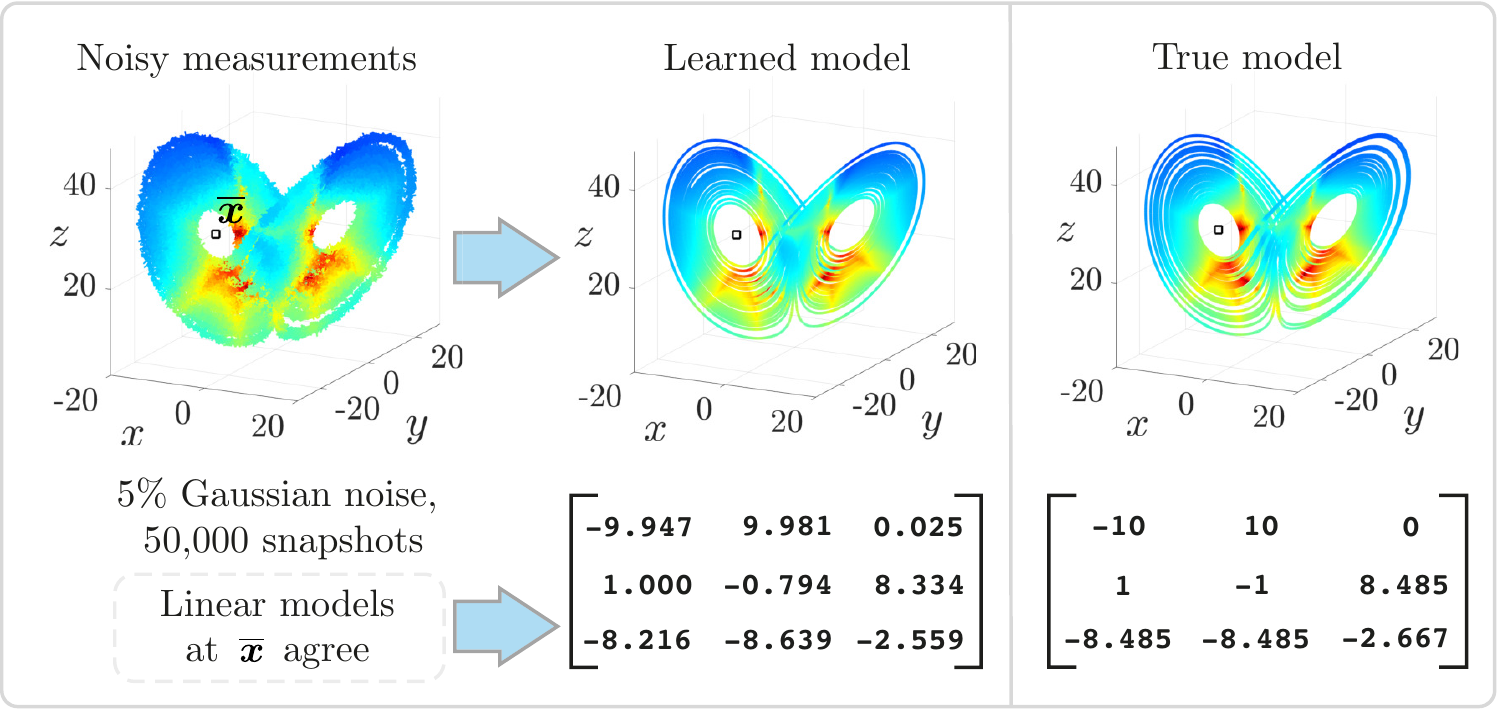}
	\caption{
	Implicit learning of the Lorenz system in the presence of noise.
	Initial measurements are corrupted by $5\%$ Gaussian noise.
	The velocity vector $\dot{\bx}$ is computed by differentiating the data measurements using total variation regularised differentiation~\cite{Chartrand2011}.
	The reconstruction captures the qualitative features of the original Lorenz system, and also accurately reproduces the local linear model at $\overline{\bx}$.
	The trajectories are colored by the adaptive time step, with red indicating a smaller time step.
	}
	\label{Fig:noise}
\end{figure*}

All machine learning algorithms must be understood in the context of their sensitivity to noise.
To explore the effects of noise, we applied our learning framework to noise-contaminated data generated by the Lorenz system from section \ref{Sec:Lorenz}.
The data are contaminated with Gaussian noise of magnitude $5\%$ of the variance of the original data; the noisy training data is visualised in the left panel of figure \ref{Fig:noise}.
We use the same parameters as section \ref{Sec:Lorenz}, but the system is now integrated to $t=50$.
Unlike section \ref{Sec:Lorenz}, we assume that we only have access to snapshot measurements of $\bx$ and velocity measurements $\dot{\bx}$ are unavailable.
Therefore, we approximate the derivative $\dot{\bx}$ from the noisy snapshot data with a total-variation regularisation scheme \cite{Chartrand2011}.
Then, we use the algorithm of section \ref{Sec:method} to learn the Lorenz system with a quadratic kernel.
The results of the learned model are illustrated in the middle panel of figure \ref{Fig:noise},
along with the true local linear model, both evaluated at the equilibrium point $\overline{\bx} = \begin{bmatrix} -\sqrt{\beta(\rho-1)} & - \sqrt{\beta(\rho-1)} & \rho - 1 \end{bmatrix}^T $.
The reconstructed trajectory shows good qualitative agreement with the true model, and the 
local linear model is a good approximation to the true linearisation.
The accuracy of these approximations usually improves as more samples are added.

\begin{figure*}
    \centering
    \includegraphics[width = \linewidth]{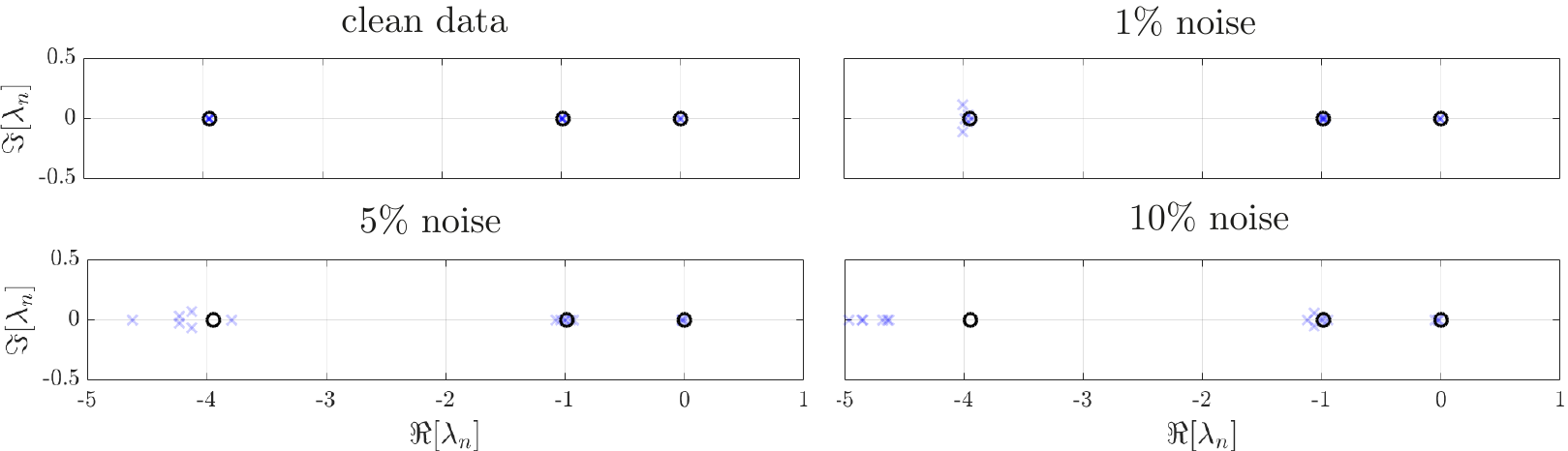}
    \caption{Identifying the spectrum of Burgers equation in the presence of noise over 20 trials. 
    The first three eigenvalues are plotted in black circles $\circ$ and their approximations are plotted in blue crosses $\times$.}
    \label{Fig:burgersNoise}
\end{figure*}

We also demonstrate the effect of noise on identifying the spectrum of the viscous Burgers' equation from section \ref{Sec:Burgers}.
Here, the kinematic viscosity is $\nu = 0.1$ and the equations are integrated to $t = 4$.
The data are snapshots of the solution which are then corrupted by Gaussian noise of varying magnitude.
No velocity measurements $\dot{\bx}$ are used and we do not de-noise the data.
Figure \ref{Fig:burgersNoise} plots the first three (repeated) eigenvalues learned by the algorithm for 20 trials with a quadratic kernel.
The first two eigenvalues are recovered accurately for small values of noise but the sensitivity to noise increases for eigenvalues of larger magnitude.
Again, these approximations can be improved by adding more samples, or by a suitable de-noising.

Further experiments indicate that, in the absence of de-noising, the algorithm is robust to noise in $\bY$ but relatively sensitive to noise in $\bX$.
These observations can be explained through the linear algebra and statistics underlying our kernel learning framework.
The impact of noise in $\bX$ is felt in two forms.
Firstly, noise may cause elements to be included in the dictionary which may otherwise have been excluded.
The dictionary is independent of $\bY$ and is therefore unaffected by noise in $\bY$.
If the sparsification parameter is not chosen carefully then noise in $\bX$ may cause the dictionary to become dense.
Secondly, noise causes errors in the final regression, whether performed online or in batch.
As is typical of least-squares regressions, our algorithm is an unbiased estimator when the noise is restricted to $\bY$.
This is because least-squares methods implicitly assume that there are no ``errors in variables'' \citep{Golub2013}.
This assumption becomes invalid when $\bX$ is contaminated by noise.
As such, when noise is present, the na\"ive pseudoinverse \eqref{Eq:kernSol} solution may prove insufficient.
For example, a similar issue arise in DMD when there is noise in $\bX$, and several approaches have been proposed to mitigate the effects of noise~\cite{Bagheri2014pof,Dawson2016,Hemati2017tcfd,Askham2018,azencot2019consistent,Scherl2020prf}, for example solving the regression problem with total least squares (TLS)~\citep{Hemati2017tcfd}.
Experiments with TLS in our present setting were found to be unsuccessful,
in part because the TLS problem is unstable \citep{Golub1980} but also because our problem is nonlinear.
In particular, TLS only guarantees the best solution to $\argmin_{\bA}\|\bA \bX - \bY \|_F$ only when the errors in $\bX$ and $\bY$ are column-wise independent and identically distributed with zero mean and covariance matrix~$\sigma^2 \bI$~
\cite[chapter 8]{VanHuffel1991}. 
Therefore, there are no guarantees of the effectiveness of TLS for our problem, which is $\argmin_{\tilde{\bW}}\|\tilde{\bW}\, k(\tilde{\bX},\bX) - \bY \|_F$. 
In the case of nonlinear dynamics, whether with SINDy or our kernel approach, the noise in $\bX$ is stretched and transformed through the nonlinearity, adding nonlinear correlations.  

In summary, we have demonstrated that our algorithm can remain effective in the presence of noise.
The Lorenz example in figure \ref{Fig:noise} indicates that the algorithm is effective when applied to derivatives computed from noisy data by total-variation regularisation.
Additionally, experiments and theory suggest that the present framework is insensitive to noise in $\bY$ but moderately sensitive to noise in $\bX$.
In addition to total-variation regularisation, there are several other methods that could be deployed to limit the effects of noise.
For example, one technique could combine the KRLS algorithm with a Kalman filter, as explored in \cite{Liu2009}.
Another filtering approach would be to use the optimal hard threshold criterion for singular values of \cite{Gavish2014}.
Fully addressing the challenge of noise is an area of active ongoing research and will be the focus of future work.

%%%%%%%%%%
%% BIBLIOGRAPHY
%%%%%%%%%%
 \small{
 \setlength{\bibsep}{5pt}
\bibliographystyle{ieeetr}
\bibliography{library}
 }
\end{document}